\documentclass[a4paper,fleqn,usenatbib,useAMS]{mnras}

\usepackage[T1]{fontenc}
\usepackage{ae,aecompl}
\usepackage{float}
\usepackage{xcolor}

\usepackage{graphics,graphicx}	
\usepackage{amsmath}	
\usepackage{amssymb}	

\usepackage{newtxtext,newtxmath}

\hypersetup{draft}

\title[Dipper-like variability of V555 Ori]{Dipper-like variability of the Gaia alerted young star V555 Ori}

\author[Z. Nagy et al.]{
Zs\'ofia Nagy,$^{1}$\thanks{E-mail: nagy.zsofia@csfk.org}
Elza Szegedi-Elek,$^{1}$
P\'{e}ter \'{A}brah\'{a}m,$^{1,2}$
\'{A}gnes K\'{o}sp\'{a}l,$^{1,2,3}$
Attila B\'{o}di,$^{1,2,4}$
\newauthor J\'{e}r\^{o}me Bouvier,$^{5}$
M\'{a}ria Kun,$^{1}$
Attila Mo\'{o}r,$^{1,2}$
Borb\'ala Cseh,$^{1}$
Anik\'o Farkas-Tak\'acs,$^{1}$
Ott\'o Hanyecz,$^{1}$ 
\newauthor Simon Hodgkin,$^{6}$
Bernadett Ign\'acz,$^{1}$ 
Csaba Kiss,$^{1,2}$
R\'eka K\"onyves-T\'oth,$^{1}$
Levente Kriskovics,$^{1,2}$ 
\newauthor G\'abor Marton,$^{1,2}$
L\'aszl\'o M\'esz\'aros,$^{1}$
Andr\'as Ordasi,$^{1}$
Andr\'as P\'al,$^{1}$
Paula Sarkis,$^{3}$
Kriszti\'an S\'arneczky,$^{1}$
\newauthor \'Ad\'am S\'odor,$^{1}$
L\'aszl\'o Szabados,$^{1}$
Zs\'ofia Marianna Szab\'o,$^{1,7}$
R\'obert Szak\'ats,$^{1}$
D\'ora Tarczay-Neh\'ez,$^{1,4}$ 
\newauthor Kriszti\'an Vida,$^{1}$
Gabriella Zsidi$^{1}$ \\
$^{1}$Konkoly Observatory, Research Centre for Astronomy and Earth Sciences, E\"otv\"os Lor\'and Research Network (ELKH), \\ H-1121 Budapest, Konkoly Thege Mikl\'os \'ut 15-17., Hungary \\
$^{2}$ELTE E\"otv\"os Lor\'and University, Institute of Physics, P\'azm\'any P\'eter s\'et\'any 1/A, H-1117 Budapest, Hungary\\
$^{3}$Max Planck Institute for Astronomy, K\"onigstuhl 17, D-69117 Heidelberg, Germany \\
$^{4}$MTA CSFK Lend\"ulet Near-Field Cosmology Research Group \\
$^{5}$Univ. Grenoble Alpes, CNRS, IPAG, F-38000 Grenoble, France\\
$^{6}$Institute of Astronomy, Madingley Road, Cambridge CB3 0HA, UK\\
$^{7}$E\"otv\"os Lor\'and University, Department of Astronomy, P\'azm\'any P\'eter s\'et\'any 1/A, H-1117 Budapest, Hungary\\
}
\date{Accepted XXX. Received YYY; in original form ZZZ}

\pubyear{2021}
 
\begin{document}
\label{firstpage}
\pagerange{\pageref{firstpage}--\pageref{lastpage}}
\maketitle

\begin{abstract}
V555 Ori is a T Tauri star, whose 1.5 mag brightening was published as a \textit{Gaia} science alert in 2017. We carried out optical and near-infrared photometric, and optical spectroscopic observations to understand the light variations. The light curves show that V555 Ori was faint before 2017, entered a high state for about a year, and returned to the faint state by mid-2018. In addition to the long-term flux evolution, quasi-periodic brightness oscillations were also evident, with a period of about 5 days. At optical wavelengths both the long-term and short-term variations exhibited colourless changes, while in the near-infrared they were consistent with changing extinction. We explain the brightness variations as the consequence of changing extinction. The object has a low accretion rate whose variation in itself would not be enough to reproduce the optical flux changes. This behaviour makes V555 Ori similar to the pre-main sequence star AA Tau, where the light changes are interpreted as periodic eclipses of the star by a rotating inner disc warp. The brightness maximum of V555 Ori was a moderately obscured ($A_V$=2.3 mag) state, while the extinction in the low state was $A_V$=6.4 mag. We found that while the \textit{Gaia} alert hinted at an accretion burst, V555 Ori is a standard dipper, similar to the prototype AA Tau. However, unlike in AA Tau, the periodic behaviour was also detectable in the faint phase, implying that the inner disc warp remained stable in both the high and low states of the system.

\end{abstract}




\section{Introduction}

Classical T Tauri stars are low-mass pre-main sequence objects surrounded by a circumstellar disc. These stars are still accreting material from the disc via magnetic field lines. Variability is a characteristic of young stars. Flux changes of a few tenth of a magnitude are typically caused either by rotating stellar spots or fluctuating accretion, producing periodic or irregular light curves, respectively.

Large amplitude brightening may also be present in the light curves when the accretion rate increases by several orders of magnitude due to thermal or gravitational instabilities in the disc, or perturbations by planets or close companions. This enhanced accretion rate causes outbursts in the optical light curve (\citealp{Herbig1977}; \citealp{HartmannKenyon1996}). Based on observations, these eruptive stars are traditionally divided into two classes: FU Orionis objects (FUors; several decades long outburst, absorption spectrum), and EX Lupi type objects (EXors, several months long repetitive outbursts, emission line dominated spectrum).

Deep fadings are also detected in light-curves of pre-main sequence variables.
The intermediate-mass UX Orionis stars (UXors) show days to weeks long eclipse-like dimming at optical wavelengths, which are likely due to discs, where dense dust clumps within an orbiting rim at the inner edge of the disc may cross the line of sight and occult the star temporarily \citep{Dullemond2003}. Similar eclipse phenomena were identified among low-mass T Tauri stars, too. The AA Tau-like variables, or dippers, show periodic dips in their light curves, caused by eclipses of an inner disc warp maintained by a tilted magnetic field (\citealp{Bouvier1999}, \citealp{Bouvier2013}). Since the brightness variations in young stars are related to the disc, infrared variability is also seen. Understanding the cause of the optical-infrared brightness variations provides unique information on the innermost part of the system.

One of the most efficient monitoring programmes to discover stars (including pre-main sequence stars) with large brightness variations is the {\it Gaia} alerts system. In 2017 January a {\it Gaia} alert was published for the pre-main sequence star V555~Ori, whose brightness increased by 1.5 mag. The source is in the Orion star forming region, at a {\it Gaia}-based distance of $388 \pm 12$~pc \citep{BailerJones2018}. Near-infrared photometric and spectroscopic measurements imply an effective temperature of $4300\pm480$~K (K4 spectral type), and a significant line-of-sight extinction of $A_V=7.98$~mag \citep{DaRio2016}. With a K4 spectral type and broad H$\alpha$ lines as a tracer of accretion (Table \ref{tab:gaussian}), it is a classical T Tauri star.

Following-up the {\it Gaia} alert, we carried out a comprehensive optical and near-infrared photometric monitoring of V555 Ori. In this paper we combine these data with archival near- and mid-infrared data to better understand the nature of V555 Ori and the nature of physical processes which affect its brightness. We explain the observations and data reduction in Sect. \ref{sect:obs}, present results on the light curves, colour variations, and the detected spectral lines in Sect. \ref{sect:results}, discuss the results in Sect. \ref{sect:discussion}, and summarise them in Sect. \ref{sect:summary}.

\section{Observations and data reduction}
\label{sect:obs}

The {\it Gaia} alert\footnote{http://gsaweb.ast.cam.ac.uk/alerts/alert/Gaia17afn/} reported a 1.5 mag brightening of V555 Ori (or Gaia17afn, RA(J2000)=05:35:08.71, Dec(J2000)=$-$04:46:52.32) on 2017 January 24. The timescale and the amplitude of the brightening suggested a potential EXor outburst.
We downloaded multi-epoch {\it Gaia} G-band photometry from the alerts service webpage. 
Figure \ref{lightcurve_gaia_wise} shows the 
{\it Gaia} light curve, covering with irregular cadence the time interval between 2015 February and 2020 October.
We supplemented these data with measurements available in public databases, and with our own new observations.

\subsection{Follow-up optical photometry}

Photometric observations in the $BVR_{\rm C}I_{\rm C}$ bands spanning the time interval between 2017 January and 2020 January were performed. We obtained optical imaging observations using the 60/90/180 cm Schmidt telescope at Konkoly Observatory (Hungary), starting from about a week after the {\it Gaia} alert. The Schmidt telescope is equipped with a 4096 $\times$ 4096 pixel Apogee Alta U16 CCD camera (pixel scale: 1.03$''$).
Images were obtained in blocks of three frames per filter, and were reduced in IRAF.\footnote{IRAF is distributed by the National Optical Astronomy Observatories, which are operated by the Association of Universities for Research in Astronomy, Inc., under cooperative agreement with the National Science Foundation.} After bias-, dark subtraction, and flat-fielding, aperture photometry was performed on each image using the IRAF 'phot' task. The instrumental magnitudes were transformed into the standard system using the Fourth U.S. Naval Observatory CCD Astrograph Catalog (UCAC4) \citep{ucac4}. The photometric errors were derived from quadratic sums of the formal errors of the instrumental magnitudes and the coefficients of the transformation equations. The results are shown in Table~\ref{schmidt_magnitudes}.

V555~Ori was also observed in the $V$, $R_{\rm{C}}$, and $I_{\rm{C}}$ bands with A Novel Dual Imaging CAMera (ANDICAM) installed on the SMARTS 1.3 m telescope at Cerro-Tololo Inter-American Observatory (CTIO), in Chile. The CCD for the ANDICAM is a Fairchild 447 2k $\times$ 2k chip. Bias and flat correction for the optical images were made by the Yale SMARTS team. The measured $V$, $R_{\rm{C}}$, and $I_{\rm{C}}$ brightnesses are presented in Table~\ref{phot_tab}.
The photometric errors were derived from quadratic sums of the formal errors of the instrumental magnitudes of the comparison stars and those of the coefficients of the transformation equations.

\subsection{Follow-up infrared photometry}

Near-infrared, $JHK_S$ measurements of V555~Ori were obtained using the ANDICAM dual-channel imager on SMARTS. The infrared (IR) detector was a Rockwell 1k$\times$1k HgCdTe ``Hawaii'' array, used with 2$\times$2 binning, with $0\farcs276$/pixel binned scale and $2\farcm4\times2\farcm$4 field of view. We used the CIT/CTIO $JHK_S$ IR filters. A five-point dithering was made to enable bad pixel removal and sky subtraction in the IR images. Calibration was performed using the 2MASS magnitudes of 2--3 bright stars in the field of view. The typical photometric accuracy was about 0.02--0.03 mag in all three bands. These numbers were computed as the error of the mean of the distribution of the individual calibration factors, derived for the different comparison stars at different dither positions. The results are presented in Table~\ref{phot_tab}.

\begin{figure*}
\includegraphics[width=\textwidth]{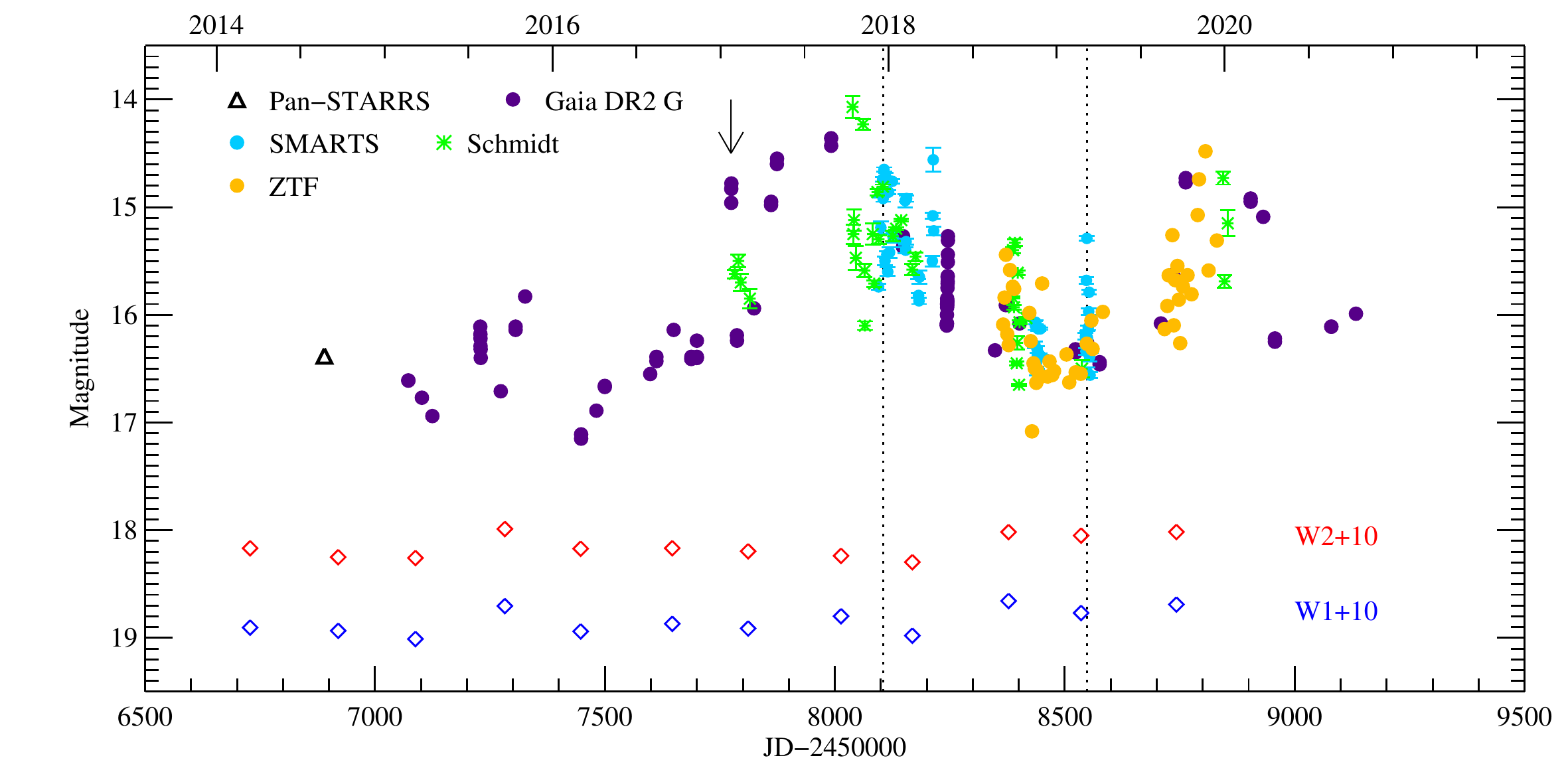}
\caption{A combined $R$-band optical light curve of V555~Ori between 2014 and 2020. Purple circles show magnitudes obtained in the {\it Gaia} $G$ filter, whose spectral response is close to that of the Cousins $R_C$-band. The black triangle shows an average $R_{\rm{C}}$ value obtained from the Pan-STARRS survey, after converting from $r$ and $g$ magnitudes. Light blue and green symbols show the $R_{\rm{C}}$ data points measured with the SMARTS and Schmidt telescopes, respectively. Data points from the ZTF survey displayed with orange symbols were converted to $R_{\rm{C}}$ magnitudes from the original $g$ and $r$ values. To show the behaviour of the source at mid-infrared wavelengths, we overplotted the WISE data with blue and red diamonds. These data points are shifted as marked in the figure. Vertical dashed lines mark the epochs when the optical spectra were taken. The arrow marks the {\it Gaia} alert.}
\label{lightcurve_gaia_wise}
\end{figure*}

\subsection{Publicly available photometric observations}

To reconstruct the light curve of V555 Ori before and after the {\it Gaia} alert we supplemented our data with photometric data available in public databases.

We used data from the Zwicky Transient Facility (ZTF), which is a time-domain survey that had first light at Palomar Observatory in 2017. ZTF uses a camera with a 47 square degree field of view mounted on the Samuel Oschin 48-inch Schmidt telescope. 
We downloaded $g$ and $r$ magnitude series of V555 Ori from the ZTF archive.\footnote{https://irsa.ipac.caltech.edu/Missions/ztf.html}

V555 Ori was observed by the Transiting Exoplanet Survey Satellite (TESS, \citealp{Ricker2015}), for 23 days between 2018 mid-December and 2019 early January, as a part of the Sector 6 observing run. The uninterrupted light curve had a cadence of 30 minutes. We retrieved the full-frame images from the Mikulski Archive for Space Telescopes (MAST) and analysed them using a FITSH-based pipeline \citep{Pal2012} providing convolution-based differential imaging algorithms and subsequent photometry on the residual images. Because the spectral sensitivity of the TESS detectors are close to the $I_{\rm{C}}$ band filter, we used our contemporaneous Schmidt $I_{\rm{C}}$ band data for the absolute calibration of the TESS photometry.

We used near-infrared JHK observations of V555~Ori obtained in 2000 March with the 2MASS telescope in Cerro Tololo. While the average value and the dispersion of the photometric measurements were published in \citet{carpenter}, we could also obtain the single epoch data points (J. Carpenter, private communication).

V555~Ori was monitored with the \textit{Spitzer} space telescope in the IRAC1 (3.6$\mu$m) and IRAC2 (4.5$\mu$m) bands, in the framework of the YSOVAR programme \citep{MoralesCalderon2011}. The observations were performed between 2009 Oct 23 and Dec 1, in the post-cryogenic phase of the mission.

NASA's Wide-field Infrared Survey Explorer (WISE, \citealp{wise_ref}) mapped the whole sky at 3.4, 4.6, 12, and 22 $\micron$ (W1, W2, W3, W4) in 2010 with an angular resolution of 6.1$\arcsec$, 6.4$\arcsec$, 6.5$\arcsec$, and 12.0$\arcsec$, respectively. 
After the depletion of the frozen hydrogen the survey continued as NEOWISE, later as NEOWISE-Reactivated period using the two shortest wavelength detectors. For each WISE epoch, we downloaded all time resolved observations from the AllWISE Multiepoch Photometry Table and from the NEOWISE-R Single Exposure Source Table in the W1 (3.4 $\micron$) and W2 (4.6 $\micron$) photometric bands, and computed their seasonal average after removing outlier data points.

\subsection{Optical spectroscopy}
\label{sec:feros}

We observed V555 Ori using the Fiber-fed Extended Range Optical Spectrograph (FEROS) mounted at MPG/ESO 2.2m telescope at the European Southern Observatory (ESO) in La Silla, Chile. FEROS is a cross-dispersed \'echelle spectrograph that works in a fixed configuration and delivers continuous spectral coverage in the $\sim3500-9200\,{\AA}$ wavelength range with a spectral resolution of $R=48\,000$ \citep{kaufer}. Two spectra were taken on 2017 December 17 in object--sky mode (i.e., the two fibers simultaneously recorded the science target and the sky background), and two spectra were taken on 2019 March 6 and 9 in object--calibration mode (when the two fibers simultaneously recorded the science target and the signal from a ThAr+Ne lamp). All four spectra were taken with 1800\,s exposure time.

\begin{figure*}
 \includegraphics[width=\textwidth, trim=0cm 2cm 0cm 2cm,clip=true]{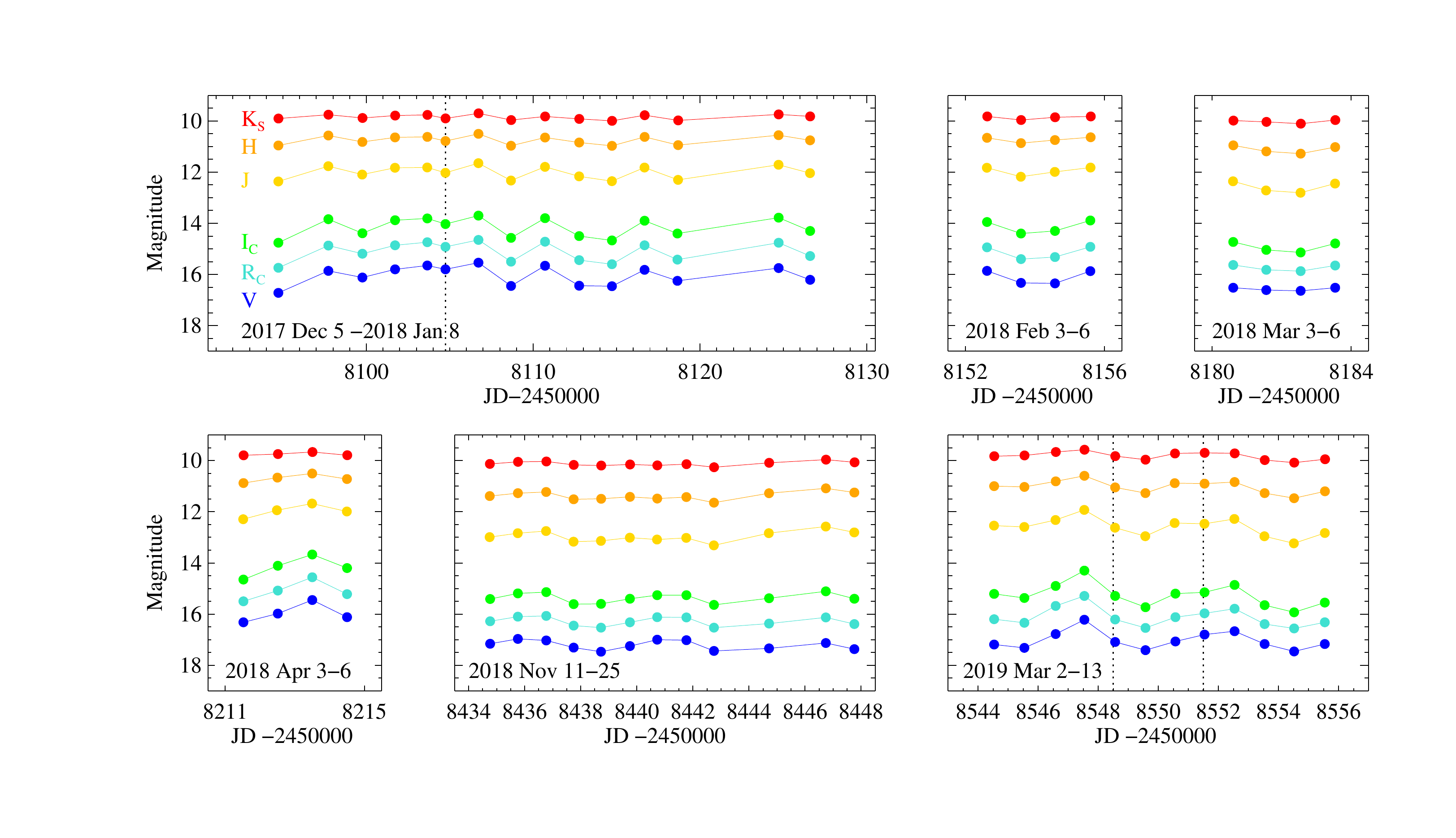}
 \caption{Subsets of the light curve measured in \textit{$V$, R$\mathrm{\sb{C}}$,  I$\mathrm{\sb{C}}$, $J$, $H$,} and \textit{K$\mathrm{\sb{S}}$} colours using the SMARTS telescope.
 The dotted lines mark the epochs of our optical spectroscopic observations.
 }
 \label{fig:all_light}
\end{figure*}

\begin{figure}
 \includegraphics[width=7.5cm]{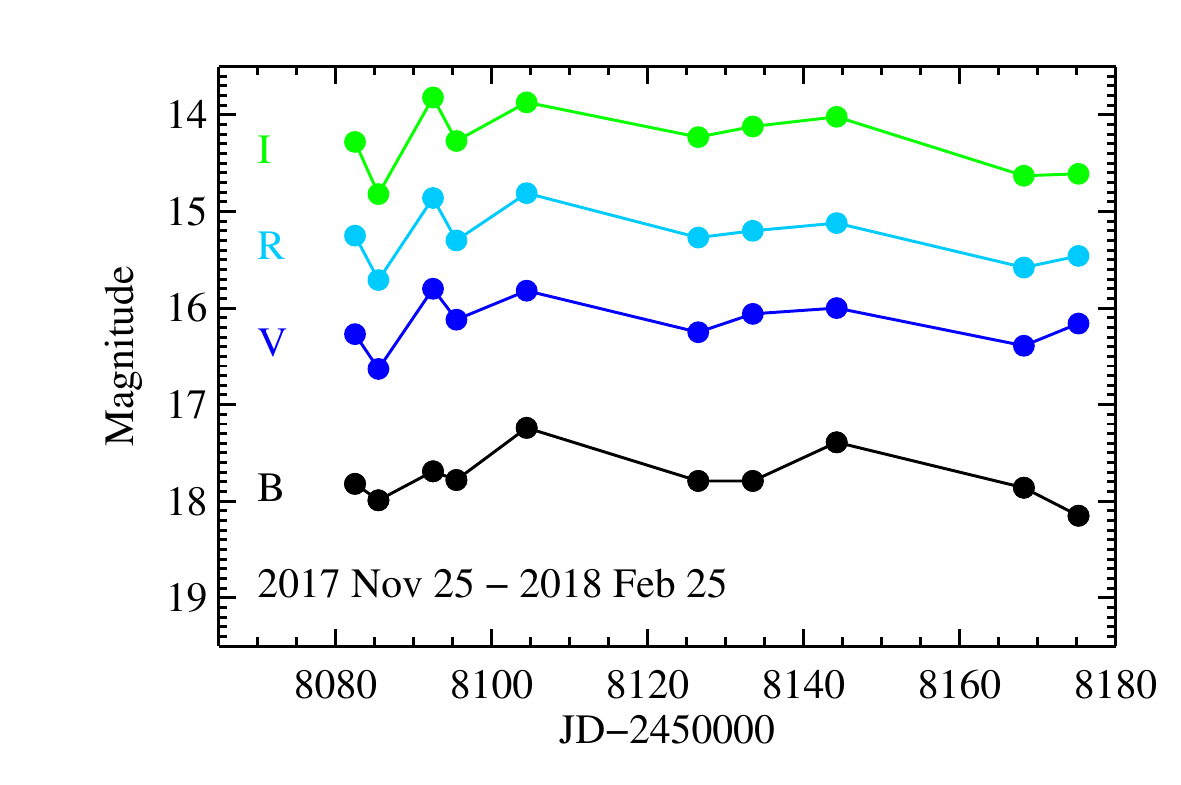}
 \includegraphics[width=7.5cm]{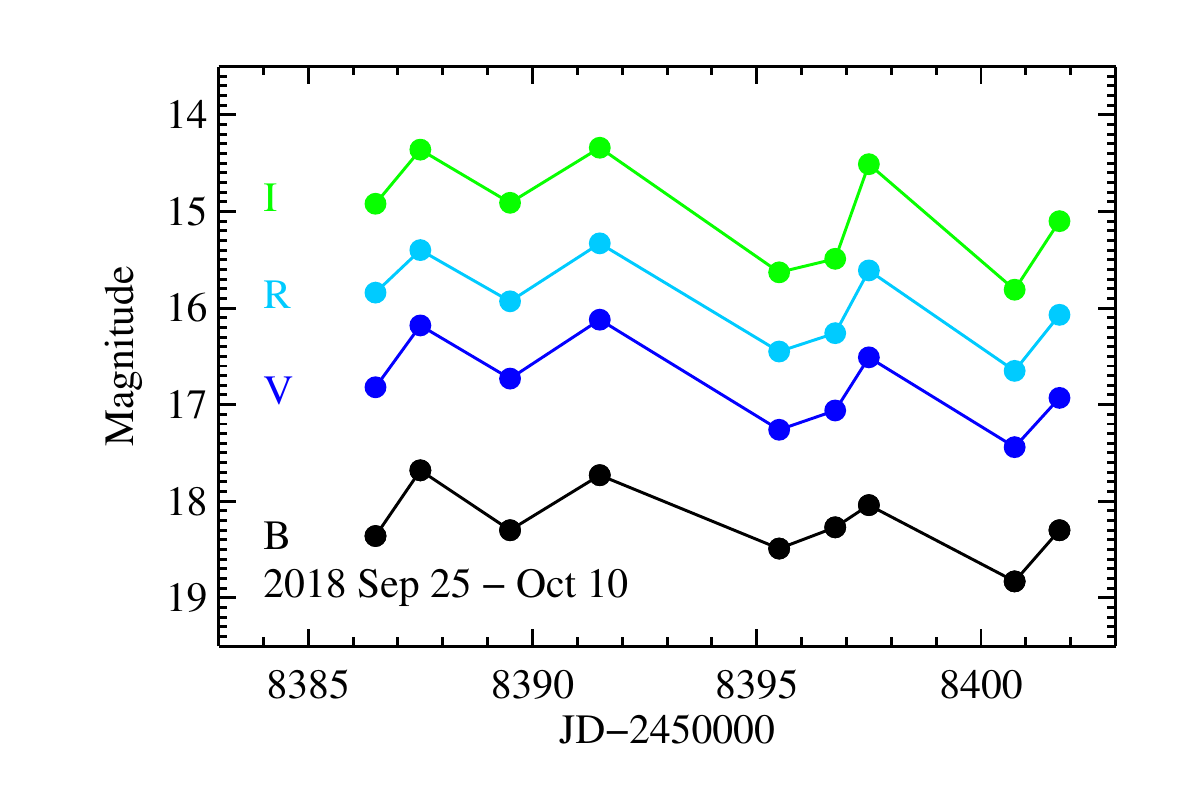}
 \caption{Subsets of the light curve measured in \textit{B, V, R$\mathrm{\sb{C}}$}, and \textit{I$\mathrm{\sb{C}}$} colours using the Schmidt telescope.}
 \label{schmidt_lightcurves}
\end{figure}

\begin{figure}
 \includegraphics[width=\columnwidth]{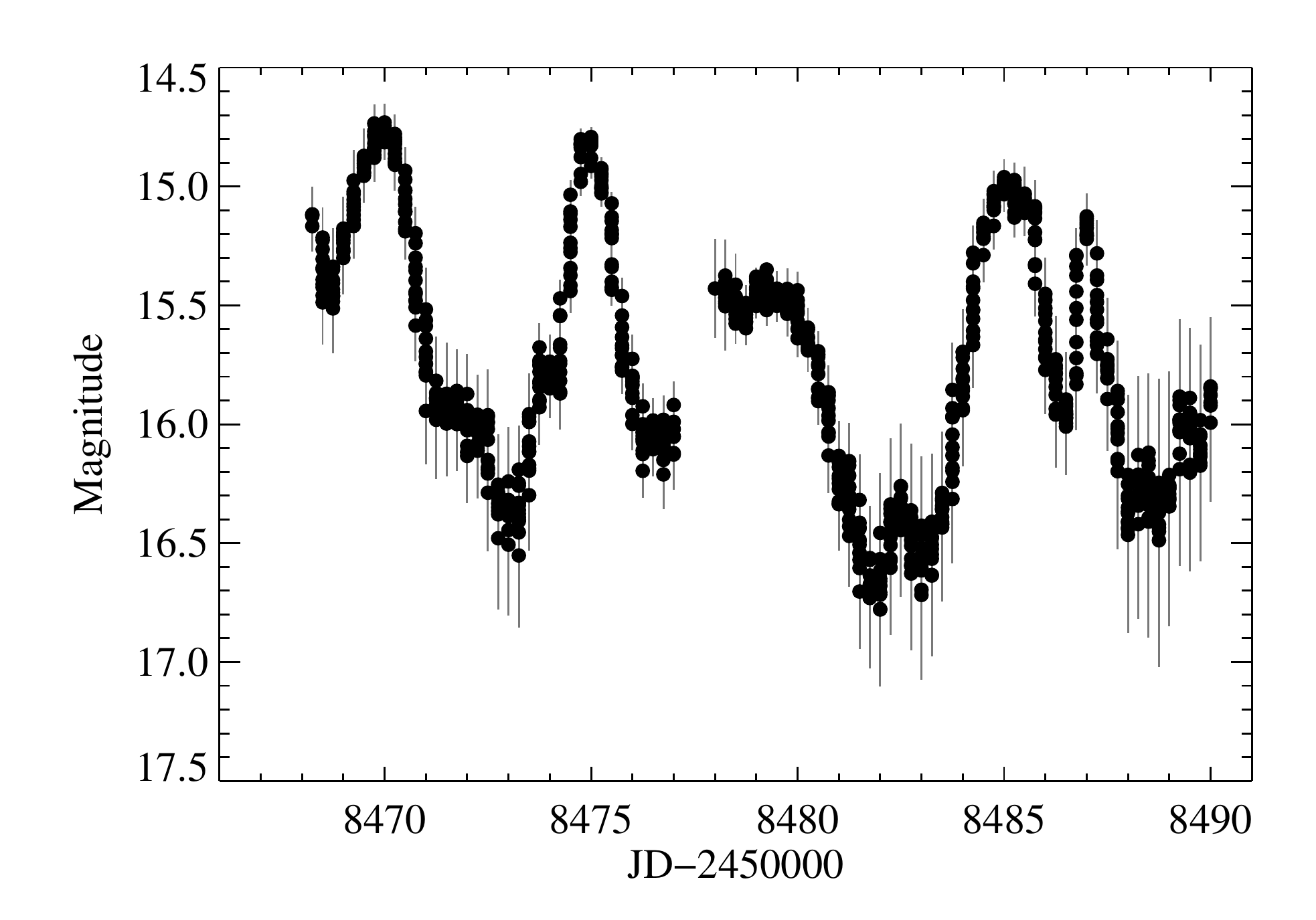}
 \caption{TESS light curve of V555 Ori.}
 \label{tess_lightcurve}
\end{figure}

We reduced the FEROS spectra using a modified version of the \textsc{ferospipe} pipeline \citep{Brahm2017} in \textsc{python}. The original pipeline has been developed to precisely measure the radial velocity of stars using cross correlation technique. To do so, it traces and extracts all the 33 available \'echelle orders, but calibrates only 25 of them, excluding eight orders that cover the $\sim6730-8230$\,{\AA} range. As this range contains several useful spectral features to study the accretion processes, we modified the code to expand the wavelength coverage. We identified the emission lines in the corresponding ThAr spectra, which are used to carry out the new wavelength calibration. The pipeline fits polynomials to the previously identified pixel value--wavelength pairs in each order to iteratively sigma clip the outliers. Then, these cleaned orders are fitted simultaneously using Chebyshev functions to improve the precision of the wavelength solution. To fit the pixel value--wavelength pairs in the newly added orders, we tested a wide range of coefficients and picked the ones where the standard deviation of the residual was approximately the same as if only the last 25 orders were used. Unfortunately, in the first 2 orders the low number of features in the ThAr spectra prevented us from a reliable line identification, thus we still had to exclude the $\sim8230-9000$\,{\AA} range.

\begin{figure*}
\includegraphics[height=6.5cm]{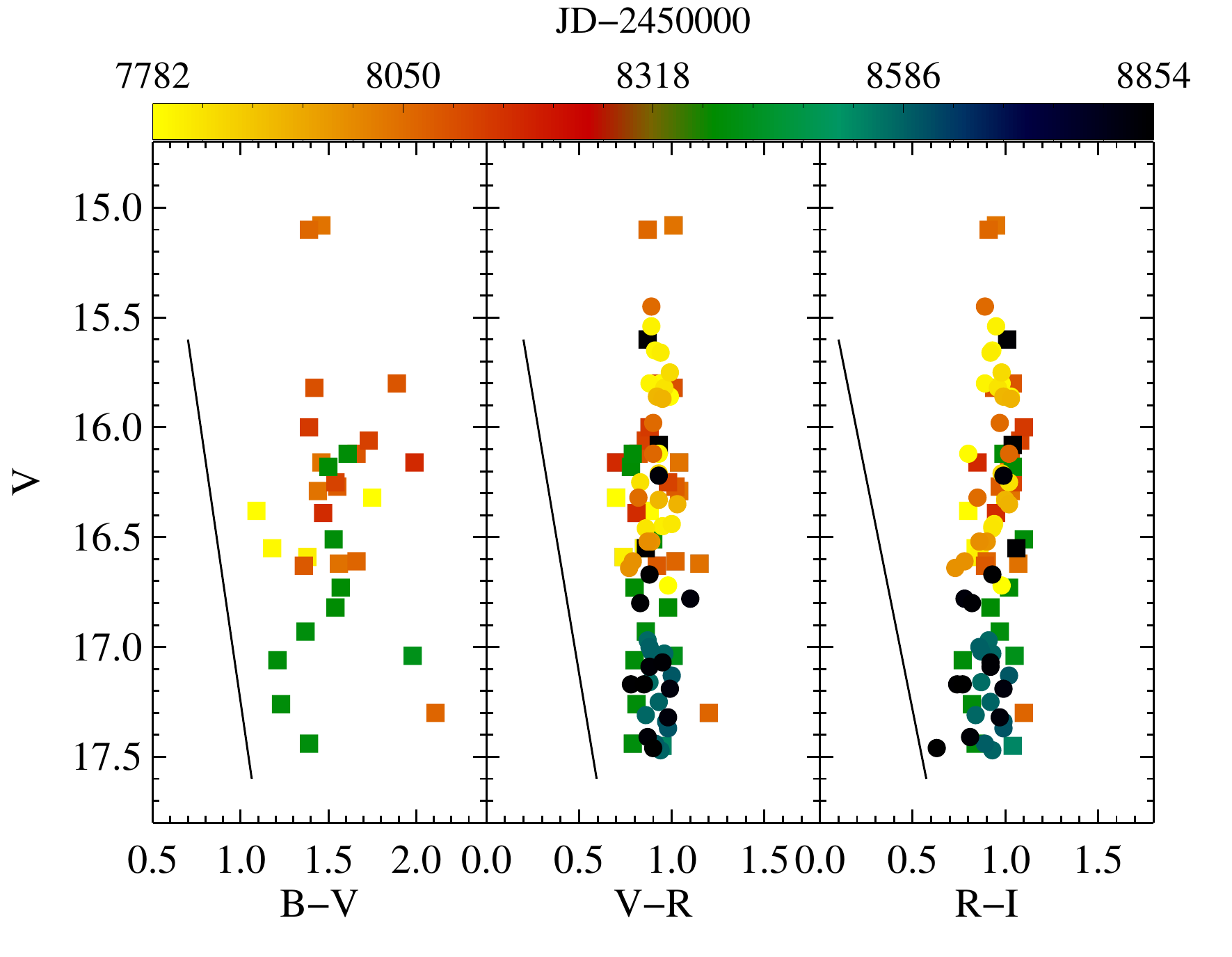}
\includegraphics[height=6.5cm, trim={0cm 0cm 0 0cm},clip]{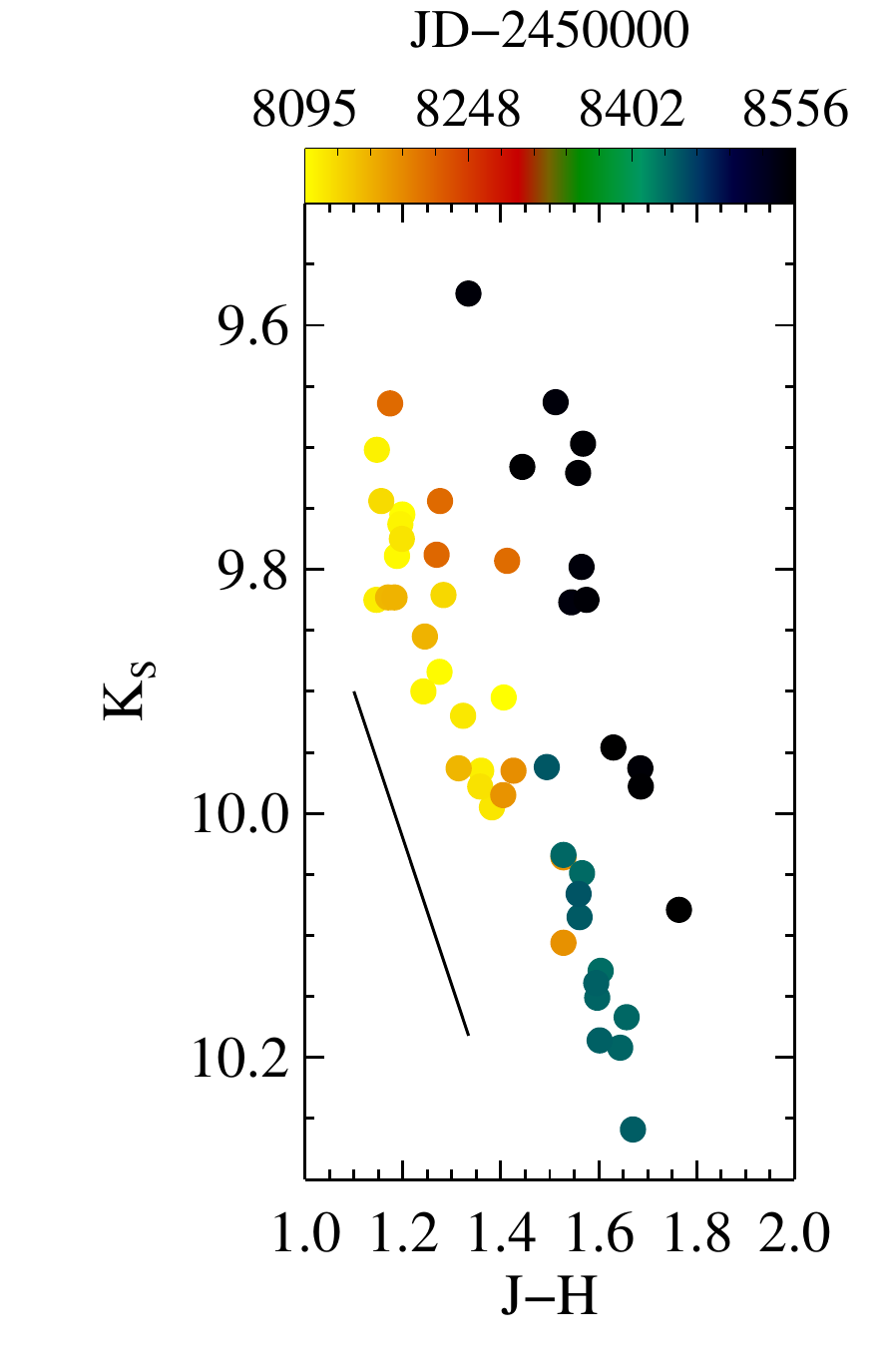}
\includegraphics[height=6.5cm,  trim={0cm 0cm 0cm 0cm},clip]{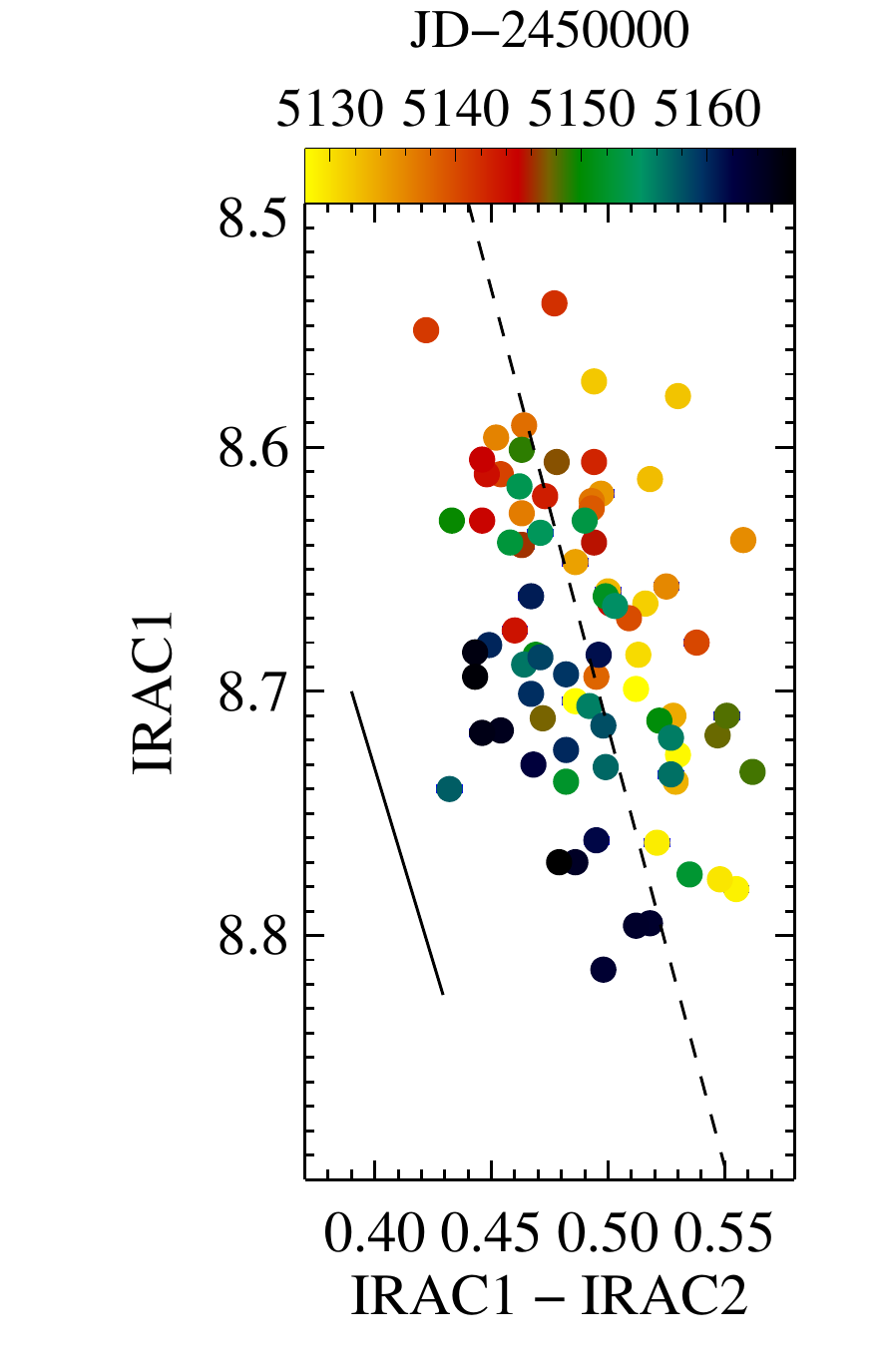}
\caption{\textit{Left and middle panels:} Colour$-$magnitude diagrams based on the optical and near-infrared data. Squares and circles show data measured with the SMARTS and Schmidt telescopes, respectively. The extinction paths correspond to $R_V=5.5$ and $A_V=2$ mag.
\textit{Right panel:} Colour$-$magnitude diagram based on the \textit{Spitzer} IRAC1 and IRAC2 colours from the YSOVAR \citep{MoralesCalderon2011} programme.
The extinction path corresponds to $R_V=5.5$ and $A_V=2$ mag. A linear fit to the data points is shown with a dashed line.
}
\label{colour2}
\end{figure*}

\begin{figure*}
\includegraphics[width=\textwidth, trim={1.1cm 0cm 0 0cm},clip]{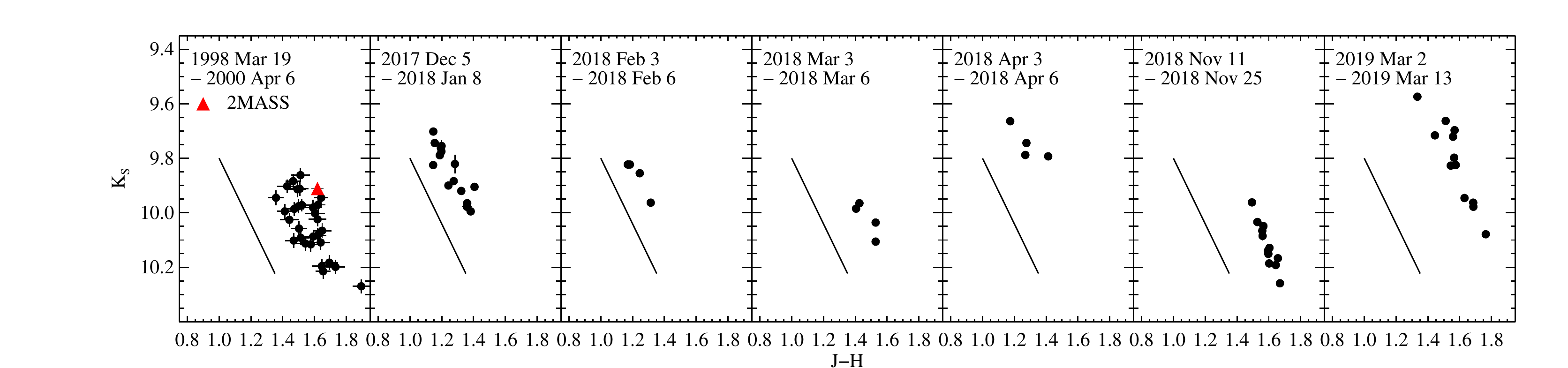}
\caption{
J--H vs K$_{\rm{S}}$ colour--magnitude diagrams and the extinction paths corresponding to $R_V=5.5$ and $A_V=3$ mag for the SMARTS data shown in Fig. \ref{fig:all_light}, and for earlier data from \citet{carpenter} and from 2MASS.
}
\label{color_magnitude_jh_k}
\end{figure*}

\section{Results}
\label{sect:results}

\subsection{Light curves}

Figure \ref{lightcurve_gaia_wise} shows a combined long-term R-band light curve of V555~Ori, using {\it Gaia} G, Pan-STARRS, Schmidt $R_{\rm{C}}$, SMARTS $R_{\rm{C}}$, and ZTF $R_{\rm{C}}$ data. The Pan-STARRS and ZTF $R_{\rm{C}}$ magnitudes were converted from the original $g$ and $r$ magnitudes, using the formulae from \citet{Tonry2012}. The optical light curve is supplemented by mid-infrared WISE photometry.

Figure \ref{lightcurve_gaia_wise} shows that already before the {\it Gaia} alert in 2017 January, V555~Ori showed optical light variations, as typical for young stars. The source brightened about one magnitude during the year preceding the alert, and stayed in a bright state for about 200 days.
During 2018, the brightness of V555 Ori showed a  decreasing trend, while continuously producing light variations.
In early 2019 a second brightening started, almost as high as in 2017, but somewhat shorter in time, and, according to the latest {\it Gaia} points, by mid 2020 V555~Ori became faint again.
Interestingly, the mid-infrared light curves do not follow the optical trends, e.g. during the optical dimming in 2018 the WISE fluxes increased. The amplitude of the infrared variability is significantly lower than in the R-band.

Some limited parts of the light curve were sampled with an almost daily cadence. Photometry in the {\it V, R$\mathrm{\sb{C}}$, I$\mathrm{\sb{C}}$, J, H, K$\mathrm{\sb{S}}$} bands within these 2--3 weeks long periods between 2017 December and 2019 March are shown in Fig. \ref{fig:all_light}. Similar plots for additional periods when only optical data were available, are displayed in Fig.~\ref{schmidt_lightcurves}. Finally, Fig.~\ref{tess_lightcurve} shows the light curve obtained by TESS in 2018 December for a period of about 3 weeks. All these figures strongly suggest short term variability on the scale of a few days. The light curves also hint at a quasi-periodic behaviour, which will be further analysed in Sect.~\ref{sec:frequency}. Remarkably, the amplitude of the short-term variations is comparable to the amplitude of the long term trends.

\subsection{Colour variations}
\label{sec:colour}

In Figure \ref{colour2} we show the colour--magnitude diagrams based on our new optical and near-infrared data, as well as on \textit{Spitzer} IRAC1 (3.6 $\mu$m) and IRAC2 (4.5 $\mu$m) archival data from 2009. The epoch of the observations is colour-coded. In the optical (left panels), while the V vs. B--V plot exhibits relatively large scatter, which is probably due to measurement uncertainties in the B-band, the V vs. V--R and V vs. R--I figures reveal invariable colours over the whole $\sim$2~mag brightness range. These constant colours do not seem to depend on the measurement date either. 

At near-IR wavelengths (middle panel) there is a linear trend in the distribution of the data points that is approximately parallel to the extinction path. However, a cloud of points is also present above the main trend. In order to investigate them, Fig.~\ref{color_magnitude_jh_k} presents separate plots for the shorter periods displayed in Fig.~\ref{fig:all_light}. In general, the variability within each period is consistent with the reddening path based on \citet{cardelli}. They seem, however, to be slightly displaced from each other by various magnitudes either in the K$_S$ brightness or in the J--H colour, suggesting that additional physical processes may also contribute to the light variability. For comparison, we also plotted earlier data from 1998--2000 (left panel), which seem to be consistent with our new observations, and imply that during the 2MASS measurement the source was in an intermediate brightness state, while later, during the campaign of \citet{carpenter}, there were epochs when the system was as faint in the K$_S$ band as the minimum state in our observations in 2018 November. The \textit{Spitzer} IRAC data are shown in Fig. \ref{colour2} (right panel). Due to the scatter seen in the data, we applied a linear fit to compare it to the extinction path. The slope of the line which fits the data is 3.6$\pm$0.9, while the slope of the extinction path at the IRAC1 and IRAC2 wavelengths is 3.2, therefore the data are also consistent with changing extinction toward the source.

The presented optical--infrared colour--magnitude diagrams revealed different trends of V555~Ori in different spectral domains: gray flux changes in the optical and extinction-like variations in the infrared. The possible explanation for this behaviour will be discussed in Sect.~\ref{sect:discussion}.

\subsection{Frequency analysis}
\label{sec:frequency}

\begin{figure}
\includegraphics[width=\columnwidth, trim={0cm 0.3cm 0cm 0.5cm},clip]{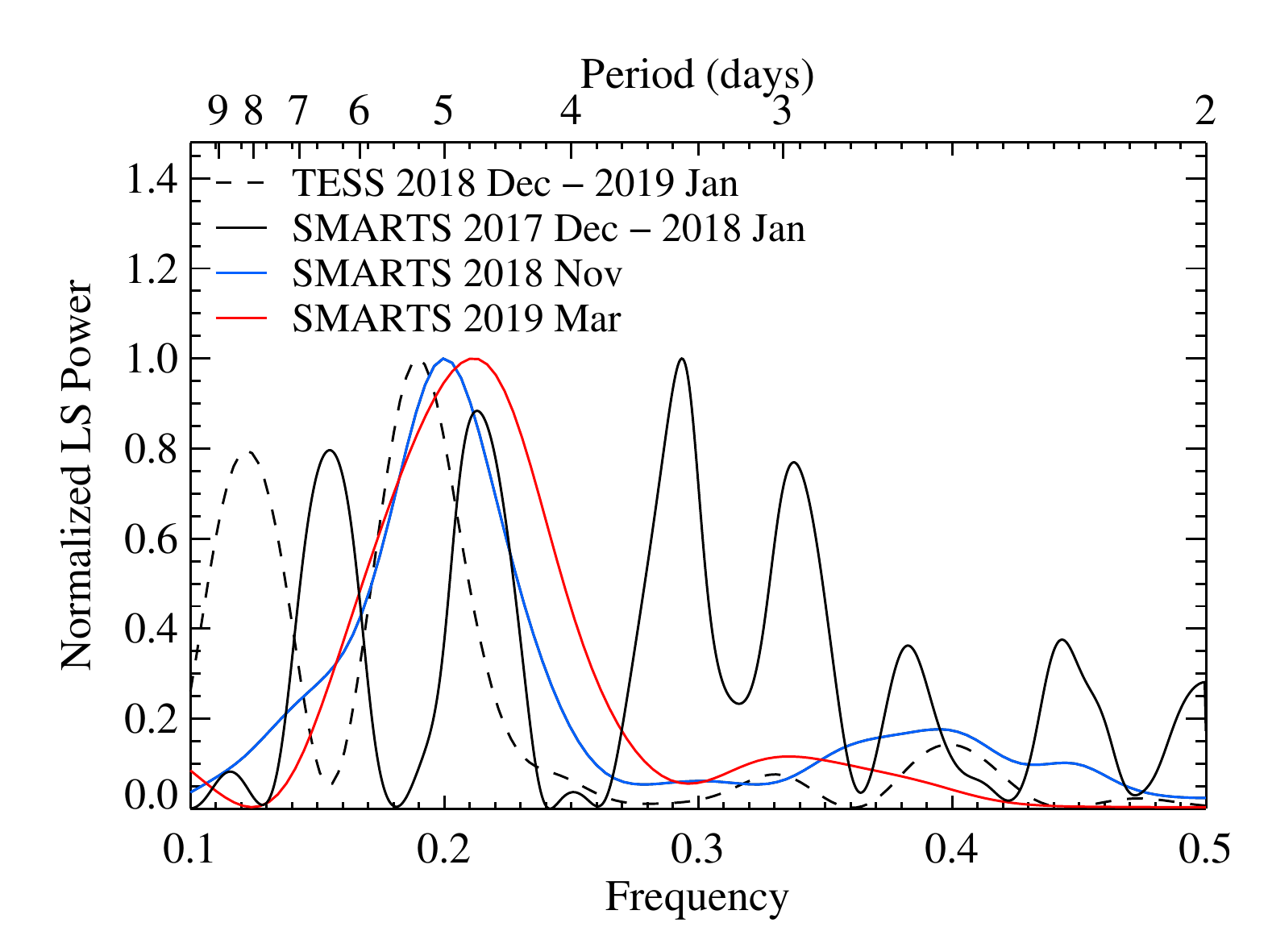}
\caption{Lomb-Scargle periodograms based on the TESS data shown in Fig. \ref{tess_lightcurve} (black solid line), and the SMARTS V-band magnitudes observed in 2017 December - 2018 January (black dashed line), November 2018 (blue line), and March 2019 (red line).}
\label{periodogram}
\end{figure}

\begin{figure}
\includegraphics[width=\columnwidth, trim={0cm 0.5cm 0cm 1.0cm},clip]{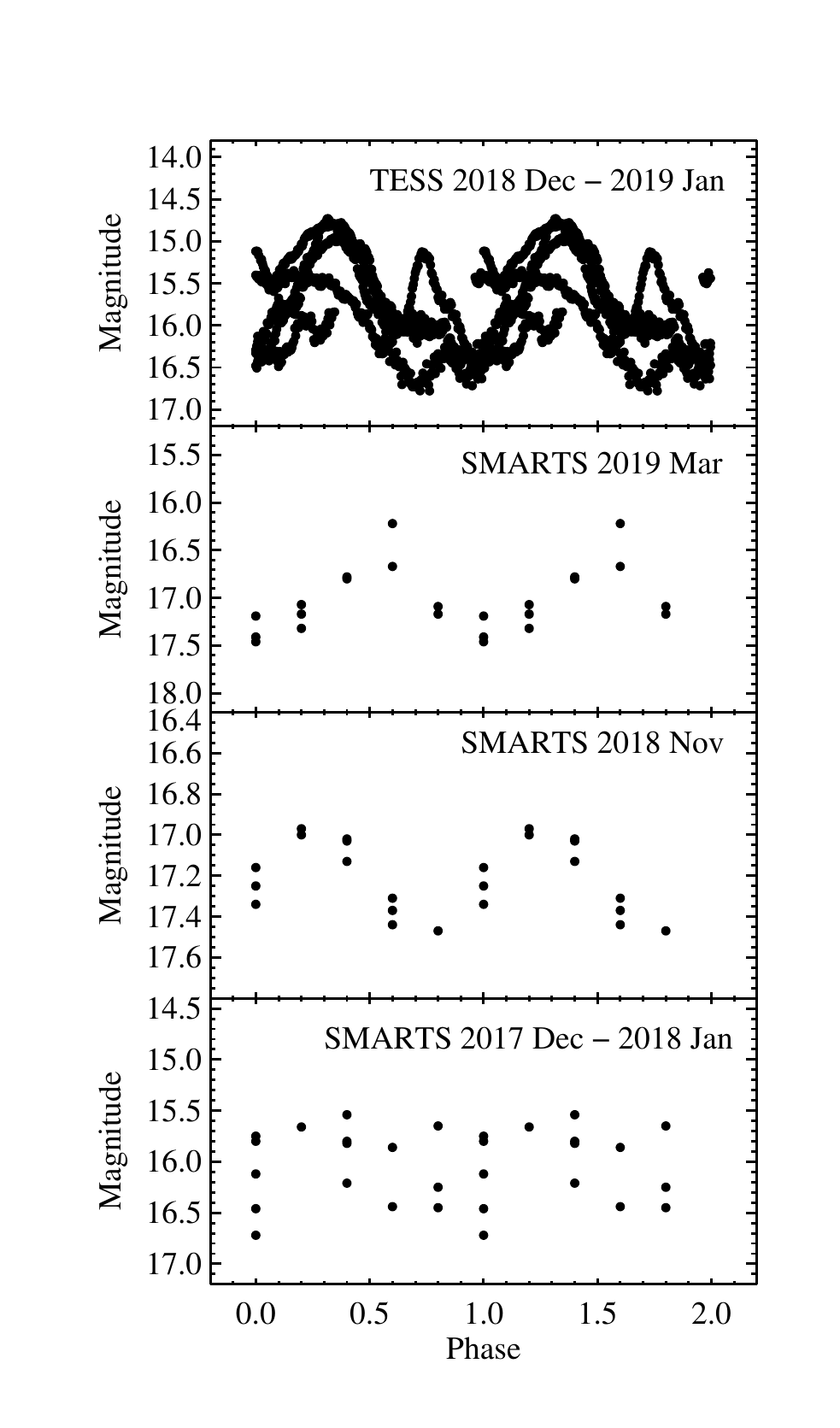}
\caption{The light curves used for the frequency analysis folded in phase with a period of 5 days.}
\label{phase_folded_lightcurves}
\end{figure}

As seen in Fig. \ref{fig:all_light}, the light curves in the different photometric bands are very similar, and they all show light variations on a time-scale of a few days. In order to characterise the virtually periodic nature of this variability, we performed a Lomb-Scargle periodogram analysis, using the public tools at the NASA Exoplanet Archive Periodogram Service\footnote{https://exoplanetarchive.ipac.caltech.edu/cgi-bin/Pgram/nph-pgram}. First we used the $V-$band photometric sequence, obtained with the SMARTS instrument and shown in Fig. \ref{fig:all_light}, to derive the most significant periods in the datasets for three epochs: 2017 Dec 5 to 2018 Jan 8; 2018 Nov 11 to 2018 Nov 25; and 2019 Mar 2 to 2019 Mar 13. Based on the periodogram results, the most significant periods in these datasets are close to 5 days, yielding 5.0$\pm$1.2 days and 4.8$\pm$1.4 days for the second and third SMARTS epochs, respectively. 
The periodogram corresponding to the first SMARTS epoch exhibits multiple periods, including a 4.7$\pm$0.3 day one. The difference between the periodogram results from first epoch compared to those from the other two may be related to the different time sampling of the data: the other two epochs were typically sampled with a daily cadence, while the 2017 Dec / 2018 Jan epoch was typically sampled with a 2-3 day cadence.
We also calculated  Lomb-Scargle periodogram for the TESS lightcurve (Fig. \ref{tess_lightcurve}). The most significant period is 5.3$\pm$0.8 days, which is close to the result obtained from the SMARTS data. 
The errors of the periods were calculated by fitting Gaussian profiles to the peaks of the periodograms, and using half of the full width at half maximum (FWHM) values of the fitted Gaussians as the errors of the periods.
The results are summarized in Fig. \ref{periodogram} and show the presence of a significant period around 5 days in both the TESS and SMARTS data. The light curves used for the frequency analysis folded in phase with a period of 5 days are shown in Fig. \ref{phase_folded_lightcurves}.

\subsection{Optical spectra}
\label{sec:spectroscopy}

The optical spectra analysed in this section represent the bright state (2017) and the faint state (2019) of V555 Ori.
As mentioned before in Sect.~\ref{sec:feros}, V555 Ori was measured in object--sky mode in 2017, but in object--calibration mode in 2019. 
The consequence of this is that the object spectra taken in 2019 are contaminated by thorium, argon, and neon lines from the calibration lamp. Therefore, when searching for emission lines in the spectra of V555 Ori, we only accepted as secure detection those lines that were present in the spectra both in 2017 and in 2019. 
Considering emission lines, we detected the H$\alpha$, H$\beta$, H$\gamma$, and H$\delta$ lines of the hydrogen Balmer series, two forbidden ionized sulphur lines ([\ion{S}{ii}] 671.8\,nm and [\ion{S}{ii}] 673.3\,nm), two forbidden ionized nitrogen lines ([\ion{N}{ii}] 655.0\,nm and [\ion{N}{ii}] 658.5\,nm), and two forbidden neutral oxygen lines ([\ion{O}{i}] 630.2\,nm and 636.5\,nm). We also detected one absorption line, that of \ion{Li}{i} at 670.8\,nm.

\begin{figure*}
 \includegraphics[height=\textwidth,angle=90]{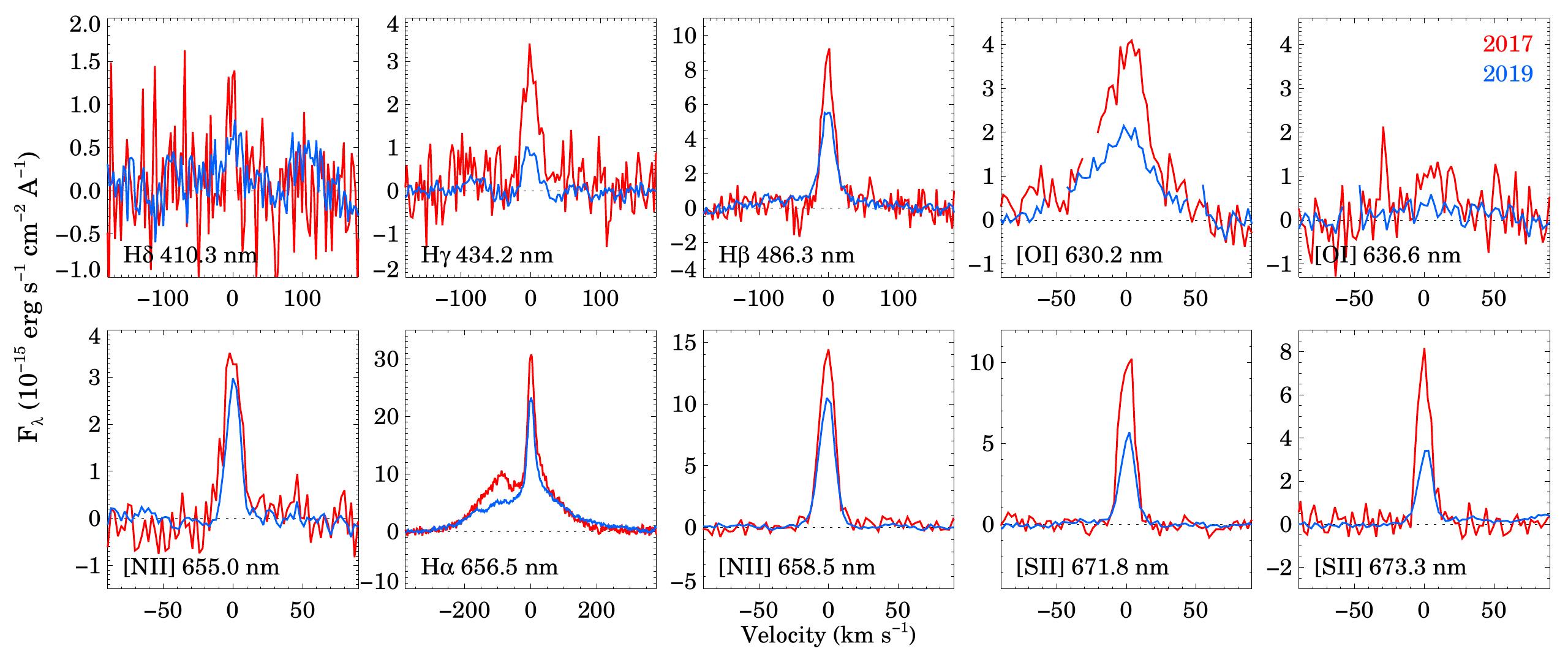}
 \caption{Emission lines detected in the FEROS optical spectra of V555 Ori in 2017 December (red) and in 2019 March (blue).}
 \label{fig:lines3}
\end{figure*}

In order to flux-calibrate the spectra, we used optical photometry taken with the SMARTS or Piszk\'estet\H{o} Schmidt telescope on the same night as the spectrum in question. For the calibration of the spectra, we estimated the continuum flux of V555 Ori at each wavelength by interpolating in wavelength between the observed broad-band photometric points. Once the spectrum was in physical units, we fitted a linear continuum to the line-free channels on both sides of each line and subtracted it. The two spectra taken with a half hour difference in 2017 and the two spectra taken with three days difference in 2019 agreed within the uncertainties, therefore we averaged the two spectra for each year to increase the signal-to-noise ratio. The calibrated, continuum-subtracted, averaged spectra for the emission lines detected in V555 Ori can be seen in Fig.~\ref{fig:lines3}.
The difference between the noise levels at the two epochs, i.e., higher signal-to-noise ratio in 2019 than in 2017, is due to different weather conditions. During the 2017 observations, the humidity was close to 60\%, the seeing was somewhat worse than 1 arcsec, while during the 2019 observations, the humidity was about 40\%, seeing slightly below 1 arcsec.

We used standard IRAF routines to measure the center location (or peak location for H$\alpha$), equivalent width, flux, and FWHM of the lines. These parameters, shown in Tab.~\ref{tab:lines}, indicate that while the equivalent widths were typically larger in 2019 than in 2017, the lines fluxes decreased from 2017 to 2019. The explanation for this is that the continuum flux also decreased from 2017 to 2019, but with a larger factor than the line fluxes did. The equivalent width of the Li line did not vary noticeably, suggesting that the veiling did not change significantly between these two epochs.

\begin{figure}
 \includegraphics[height=\columnwidth,angle=90]{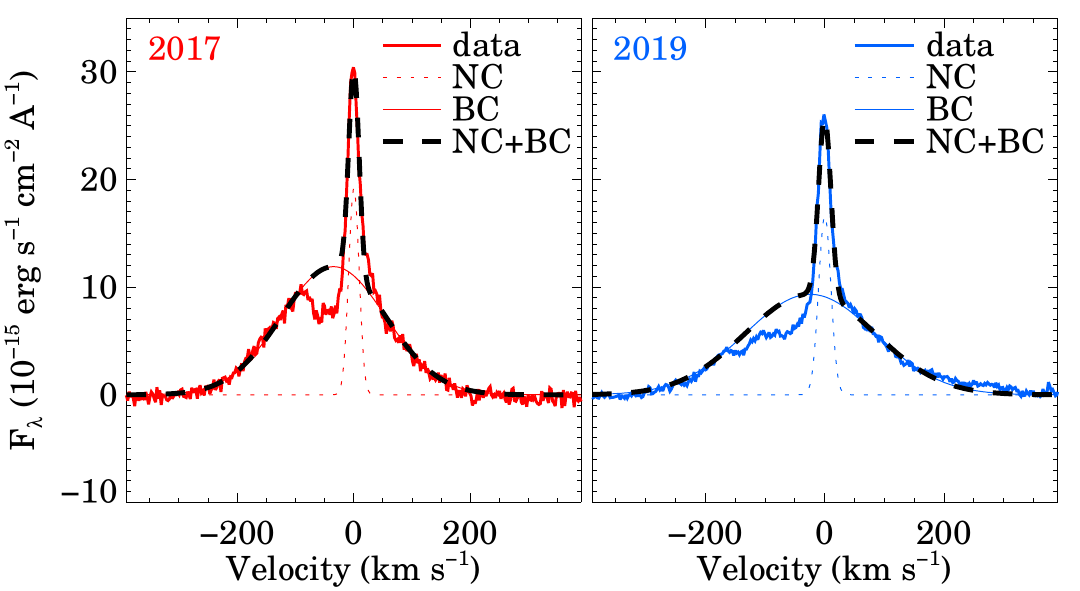}
 \caption{Profiles of the H$\alpha$ line and Gaussian fits.}
 \label{fig:halpha}
\end{figure}

Most emission lines we detected display simple Gaussian profiles. The only exception is H$\alpha$, which is more complex. In Fig.~\ref{fig:halpha} we fitted the observed H$\alpha$ profiles with the sum of two Gaussians, a narrow and a broad component (NC and BC). Tab.~\ref{tab:gaussian} shows the fitted parameters of the Gaussians in 2017 and 2019. The two Gaussians reproduce the profiles quite well, except for some extra absorption apparent between $-$85\,km\,s$^{-1}$ and $-25$\,km\,s$^{-1}$ in 2017 and between $-$150\,km\,s$^{-1}$ and $-$25\,km\,s$^{-1}$ in 2019. The Gaussian width ($\sigma$) is on the order of 20\,km\,s$^{-1}$ for the NC and 100\,km\,s$^{-1}$ for the BC. There is about 6 times more flux in the BC. While the NC appears at around the systemic velocity, the BC is blueshifted by $-$25 to $-$35\,km\,s$^{-1}$. 
The narrow component may be related to postshock gas and the broad component to the accretion flow (\citealp{Hartmann2016} and references therein).

Fig.~\ref{fig:lines3} and Tab.~\ref{tab:lines} show that the metallic emission lines became weaker in 2019 compared to 2017, but there are no discernible differences in the line profiles. We used the CHIANTI Database V 9.0\footnote{http://chiantidatabase.org/} to calculate model line ratios for our observed lines for different $n_e$ electron densities and $T_e$ electron temperatures \citep{dere1997, dere2019}. 
For the observed [\ion{O}{i}] and [\ion{N}{ii}] lines, the line ratios do not depend on $n_e$ or $T_e$, given that their upper levels are the same \citep{StoreyZeippen2000}, therefore we can simply compare the theoretical line ratio with the observed one.
The theoretical line ratio for [\ion{O}{i}] 630.2\,nm / [\ion{O}{i}] 636.5\,nm is 3.13, while the measured ratios are 5.18$\pm$0.16 and 4.77$\pm$0.78 in 2017 and in 2019, respectively. 
The theoretical line ratio for [\ion{N}{ii}] 655.0\,nm / [\ion{N}{ii}] 658.5\,nm is 0.34, while the measured ratios are 0.29$\pm$0.03 and 0.25$\pm$0.01 in 2017 and in 2019, respectively. Based on the observed ratios the lines are not saturated, i.e. they are optically thin, as otherwise the ratios would be closer to 1.
The good agreement between the theoretical and observed line ratios for [\ion{N}{ii}] also suggests that they are formed in low-density gas, because the theoretical line ratios are valid in the low-density limit, otherwise collisional de-excitation would suppress these forbidden transitions.
The discrepancy for [\ion{O}{i}] may be because of the uncertainty of the observed line ratio due to very low signal-to-noise detection of the [\ion{O}{i}] 636.5\,nm line. 
The [\ion{S}{ii}] 671.8\, nm / [\ion{S}{ii}] 673.3\,nm ratio depends both on $n_e$ and $T_e$, given that the upper levels of the two transitions are different \citep{Canto1980}. The measured ratios are 1.30$\pm$0.02 and 1.70$\pm$0.04 in 2017 and in 2019, respectively. The model values typically give ratios above one for log($n_e$)<2.5, while the ratio is below one for higher electron densities. We found no parameter combination to give a ratio larger than about 1.47. The largest ratios can be obtained for temperatures around 6000\,K. The observed ratios suggest that the [\ion{S}{ii}] lines are emitted by gas that has an electron density of 100\,cm$^{-3}$ or less and temperature probably close to 6000\,K.
These values imply low ionization fraction and low temperature compared to typical values in jets of young stellar objects ($N_{\rm{e}} \sim 10^4$ cm$^{-3}$ and $T_{\rm{e}} \sim 1.1 - 1.5 \times 10^4$ K,  \citealp{GomezdeCastroPudritz1993}). On the other hand, the values of V555 Ori are higher compared to characteristic values of photodissociation regions (few cm$^{-3}$ and $\sim$1000 K, \citealp{HollenbachTielens1997}). These may point to a weak disc wind in V555 Ori, with C-shocks rather than J-shocks \citep{Hollenbach1997}.

\begin{table}
\caption{Parameters of Gaussian components fitted to the H$\alpha$ lines of V555 Ori.}
\label{tab:gaussian}
\begin{tabular}{l@{}cc}
\hline
Parameter   & Value in 2017 & Value in 2019 \\ 
\hline
\multicolumn{3}{c}{NC} \\
\hline
Height ($10^{-15}$ erg\,s$^{-1}$\,cm$^{-2}$\,{\AA}$^{-1}$) & 19.25 $\pm$ 0.23 & 16.65 $\pm$ 0.45 \\
Position (km\,s$^{-1}$) & 0.14 $\pm$ 0.01 & $-$0.03 $\pm$ 0.01  \\
$\sigma$ (km\,s$^{-1}$) & 8.89 $\pm$ 0.05 & 10.29 $\pm$ 0.19
\\

\hline
\multicolumn{3}{c}{BC} \\
\hline
Height ($10^{-15}$ erg\,s$^{-1}$\,cm$^{-2}$\,{\AA}$^{-1}$) & 11.91 $\pm$ 0.20 & 9.34 $\pm$0.21 \\
Position (km\,s$^{-1}$) & $-$34.7 $\pm$ 0.2 & $-$24.3 $\pm$ 0.38  \\
$\sigma$ (km\,s$^{-1}$) & 90.0 $\pm$ 0.32   & 109.7 $\pm$ 1.1 \\

\hline
\end{tabular}
\end{table}

\begin{table*}
\caption{Parameters of the lines detected in V555 Ori. The FWHM values are uncorrected for the instrumental line broadening.}
\label{tab:lines}
\begin{tabular}{ccccccc}
\hline
Line           & Year & Center             & Eqv. width       & Flux                                   & Corrected flux                         & FWHM \\
               &      & (\AA)              & (\AA)            & ($10^{-15}$\,erg\,s$^{-1}$\,cm$^{-2}$) & ($10^{-15}$\,erg\,s$^{-1}$\,cm$^{-2}$) & (\AA)\\
\hline
H$\alpha$ 656.5 nm & 2017 & 6564.66 $\pm$ 0.01 & $-$39.5 $\pm$ 1.0 & 67.8 $\pm$ 0.5 & 417 $\pm$ 3   & 0.53 $\pm$ 0.01$^{a}$ \\
H$\alpha$ 656.5 nm & 2019 & 6564.64 $\pm$ 0.01 & $-$75.1 $\pm$ 3.0 & 55.7 $\pm$ 0.5 & 1030 $\pm$ 10 & 0.46 $\pm$ 0.01$^{a}$ \\

H$\beta$ 486.3 nm  & 2017 & 4862.74 $\pm$ 0.01 & $-$5.45 $\pm$ 0.33 & 3.79 $\pm$ 0.16 & 38.9 $\pm$ 1.6 & 0.38 $\pm$ 0.03 \\
H$\beta$ 486.3 nm  & 2019 & 4862.70 $\pm$ 0.01 & $-$12.7 $\pm$ 1.7  & 2.73 $\pm$ 0.13 & 76.8 $\pm$ 3.7 & 0.46 $\pm$ 0.05 \\

H$\gamma$ 434.2 nm & 2017 & 4341.73 $\pm$ 0.01 & $-$2.79 $\pm$ 0.01 & 1.07 $\pm$ 0.01 & 13.4 $\pm$ 0.1 & 0.36 $\pm$ 0.03 \\
H$\gamma$ 434.2 nm & 2019 & 4341.71 $\pm$ 0.01 & $-$2.41 $\pm$ 0.29 & 0.26 $\pm$ 0.05 &  8.6 $\pm$ 1.6 & 0.30 $\pm$ 0.02 \\

H$\delta$ 410.3 nm & 2017 & 4102.91 $\pm$ 0.01 & $-$0.45 $\pm$ 0.10 & 0.18 $\pm$ 0.06 & 2.5 $\pm$ 0.8 & 0.20 $\pm$ 0.05 \\
H$\delta$ 410.3 nm & 2019 & 4102.97 $\pm$ 0.01 & $-$1.85 $\pm$ 0.30 & 0.23 $\pm$ 0.11 & 8.1 $\pm$ 3.9 & 0.38 $\pm$ 0.10 \\

[\ion{O}{i}]  630.2 nm & 2017 & 6301.97 $\pm$ 0.02 & $-$2.26 $\pm$ 0.04 & 4.77 $\pm$ 0.09 & 31.5 $\pm$ 0.6 & 1.20 $\pm$ 0.13 \\ \relax
[\ion{O}{i}]  630.2 nm & 2019 & 6302.06 $\pm$ 0.02 & $-$3.75 $\pm$ 0.04 & 2.58 $\pm$ 0.09 & 50.6 $\pm$ 1.8 & 1.24 $\pm$ 0.10 \\ \relax
[\ion{O}{i}]  636.6 nm & 2017 & 6365.78 $\pm$ 0.02 & $-$0.50 $\pm$ 0.01 & 0.92 $\pm$ 0.01 &  6.0 $\pm$ 0.1 & 0.76 $\pm$ 0.15 \\ \relax
[\ion{O}{i}]  636.6 nm & 2019 & 6365.54 $\pm$ 0.04 & $-$1.47 $\pm$ 0.17 & 0.54 $\pm$ 0.06 & 10.4 $\pm$ 1.2 & 1.16 $\pm$ 0.25 \\

[\ion{N}{ii}] 655.0 nm & 2017 & 6549.91 $\pm$ 0.01 & $-$0.83 $\pm$ 0.03 & 1.38 $\pm$ 0.04 &  8.5 $\pm$ 0.2 & 0.31 $\pm$ 0.03 \\ \relax
[\ion{N}{ii}] 655.0 nm & 2019 & 6549.93 $\pm$ 0.01 & $-$1.67 $\pm$ 0.07 & 0.84 $\pm$ 0.01 & 15.6 $\pm$ 0.2 & 0.24 $\pm$ 0.02 \\ \relax

[\ion{N}{ii}] 658.5 nm & 2017 & 6585.31 $\pm$ 0.01 & $-$2.24 $\pm$ 0.07 & 4.61 $\pm$ 0.25 & 28.2 $\pm$ 1.5 & 0.28 $\pm$ 0.03 \\ \relax
[\ion{N}{ii}] 658.5 nm & 2019 & 6585.29 $\pm$ 0.01 & $-$4.35 $\pm$ 0.29 & 3.35 $\pm$ 0.07 & 61.5 $\pm$ 1.3 & 0.29 $\pm$ 0.02 \\ \relax

[\ion{S}{ii}] 671.8 nm & 2017 & 6718.36 $\pm$ 0.01 & $-$1.50 $\pm$ 0.04 & 3.23 $\pm$ 0.02 & 19.0 $\pm$ 0.1 & 0.28 $\pm$ 0.01 \\ \relax
[\ion{S}{ii}] 671.8 nm & 2019 & 6718.38 $\pm$ 0.01 & $-$2.63 $\pm$ 0.17 & 1.65 $\pm$ 0.05 & 29.4 $\pm$ 0.9 & 0.27 $\pm$ 0.01 \\ \relax

[\ion{S}{ii}] 673.3 nm & 2017 & 6732.73 $\pm$ 0.01 & $-$1.13 $\pm$ 0.01 & 2.48 $\pm$ 0.04 & 14.6 $\pm$ 0.2 & 0.25 $\pm$ 0.01 \\ \relax
[\ion{S}{ii}] 673.3 nm & 2019 & 6732.76 $\pm$ 0.01 & $-$1.29 $\pm$ 0.06 & 0.97 $\pm$ 0.02 & 17.2 $\pm$ 0.4 & 0.26 $\pm$ 0.01 \\ \relax

\ion{Li}{i}   671.0 nm & 2017 & 6709.95 $\pm$ 0.03 & 0.47 $\pm$ 0.04 & \dots & \dots & 0.85 $\pm$ 0.20 \\
\ion{Li}{i}   671.0 nm & 2019 & 6709.95 $\pm$ 0.02 & 0.42 $\pm$ 0.05 & \dots & \dots & 0.56 $\pm$ 0.10 \\

\hline
\multicolumn{6}{l}{$^{a}$\footnotesize{FWHM of the narrow component.}} \\
\end{tabular}
\end{table*}

\section{Discussion}
\label{sect:discussion}

\subsection{Long-term variability: obscuration or outburst?}

Fig.~\ref{lightcurve_gaia_wise} demonstrates that V555~Ori underwent long-term brightness variations, including the $\sim$1.5 magnitude brightening corresponding to the \textit{Gaia} alert in early 2017, followed by a high state of $\sim$200 days, a dimming to reach a minimum around late 2018 - early 2019, and a rise again in early 2020. These flux changes may be caused by variable circumstellar extinction and/or fluctuations in the accretion rate. In the following, we will quantify the changes in visual extinction based on the photometric data, then in the accretion rate based on our spectroscopic data.

\begin{figure}
\centering
 \includegraphics[width=8cm]{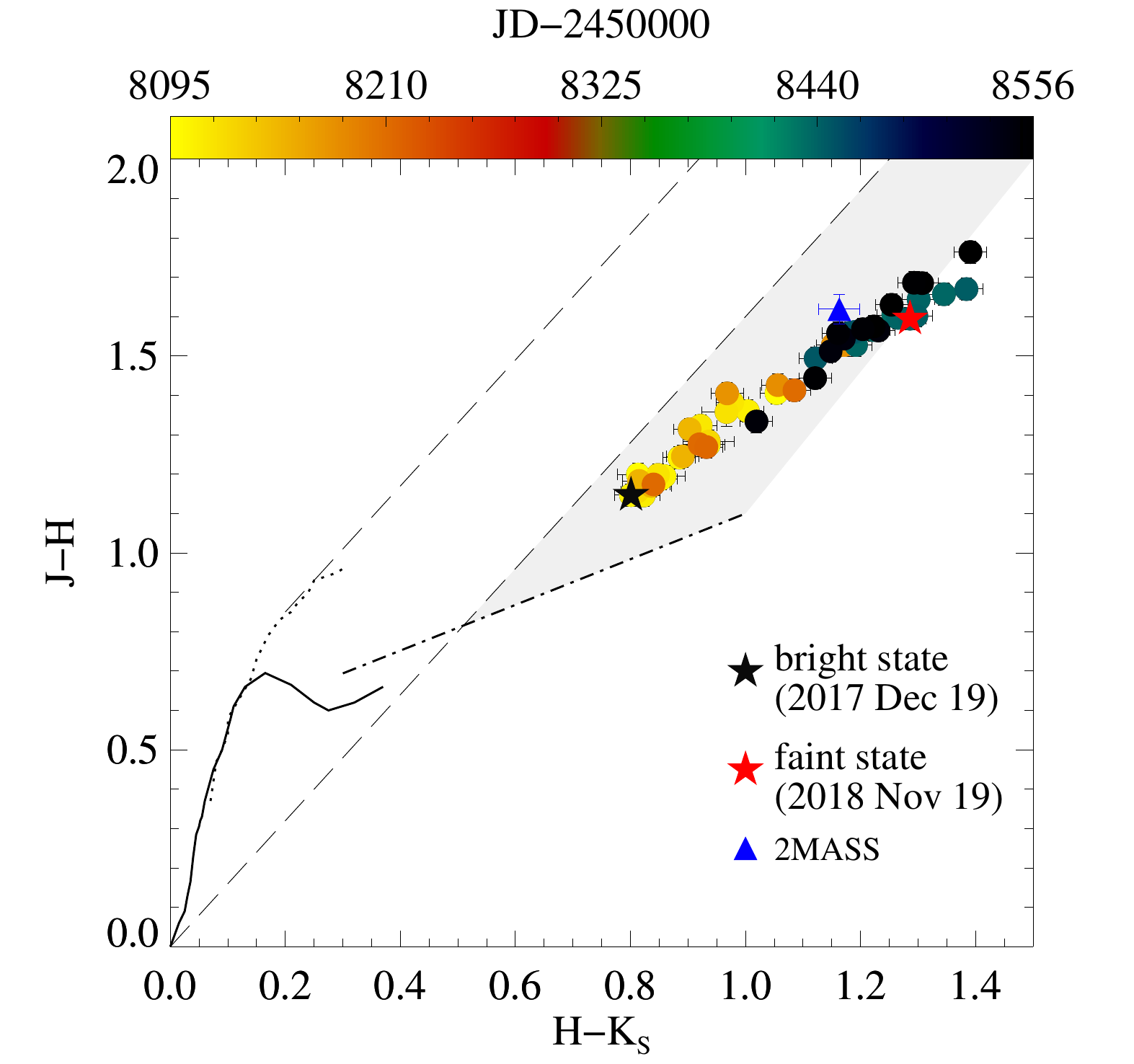}
 \caption{(J$-$H) vs. (H$-$K$\sb{\mathrm{S}}$) colour--colour diagram. The solid curve shows the colours of the zero-age main sequence, and the dotted line represents the giant branch \citep{loci}. The long-dashed lines delimit the area occupied by the reddened normal stars \citep{cardelli}. The dash-dotted line is the locus of unreddened T Tau stars \citep{meyer1997} and the gray shaded band borders the area of the reddened K$\sb{\mathrm{S}}$-excess stars. 
The coloured filled circles are SMARTS data, with their colours representing the date of the observations.}
\label{infra_colour}
\end{figure}

To measure the visual extinction in different brightness states of the long-term evolution, we analysed the location of the source in the near-infrared colour--colour diagram (Fig.~\ref{infra_colour}). In order to separate the long-term trends from short-term quasi-periodic variability, we scrutinised the light curves in Fig.~\ref{fig:all_light}, and visually identified all epochs corresponding to maximum brightness of the $\sim$5 day periodicity. We assume that at these phases no extinction component related to the short-term variability affects the system's apparent magnitude. Among these peak epochs, we found that at infrared wavelengths V555~Ori was brightest on 2017 December 19, and faintest on 2018 November 19. 

Using the expression from \citet{cardelli}, we measured the visual extinction of the source in the brightest epoch by projecting its location in Fig.~\ref{infra_colour} to the line representing the locus of unreddened T Tau stars \citep{meyer1997} along the extinction path. This method gives $A_V$=2.3 mag, which may largely be the interstellar reddening in the direction of the source. To determine the extinction in the faintest state, we measured the J--H difference magnitude between the brightest and faintest data points in Fig.~\ref{infra_colour} (black and red asterisks, respectively), and converted it to A$_V$ using the extinction law of \citet{cardelli}. We assumed an $R_V$ of 5.5, as suggested by \citet{DaRio2016} as an average value for the Orion~A cloud. Adding to this value the bright state extinction of 2.3 mag, this approach resulted in 6.4 mag for 2018 November 19. Thus, we conclude that the long-term infrared variations of V555~Ori can be explained by changing extinction along the line-of-sight, whose value varies between A$_V$=2.3~mag and 6.4~mag. The extinction-based interpretation is also supported by the the middle and right panels of Fig.~\ref{colour2}.

Note, that at optical wavelengths the flux changes do not follow what would be expected from changing obscuration, whose possible explanations will be discussed in Sect.~\ref{sec:shortterm}. Our A$_V$ value of 6.4~mag, obtained for the faintest state, is not far from the $\sim$8.0 mag derived by \citet{DaRio2016} for V555 Ori using the (J--H) colour.
In Fig.~\ref{infra_colour}, V555 Ori appears as a moderately reddened T Tau star, whereas in the faint state its near-infrared colours move to the right, beyond the area occupied by reddened Class II young stellar objects \citep{meyer1997}. 
 
\begin{figure}
 \includegraphics[width=\columnwidth]{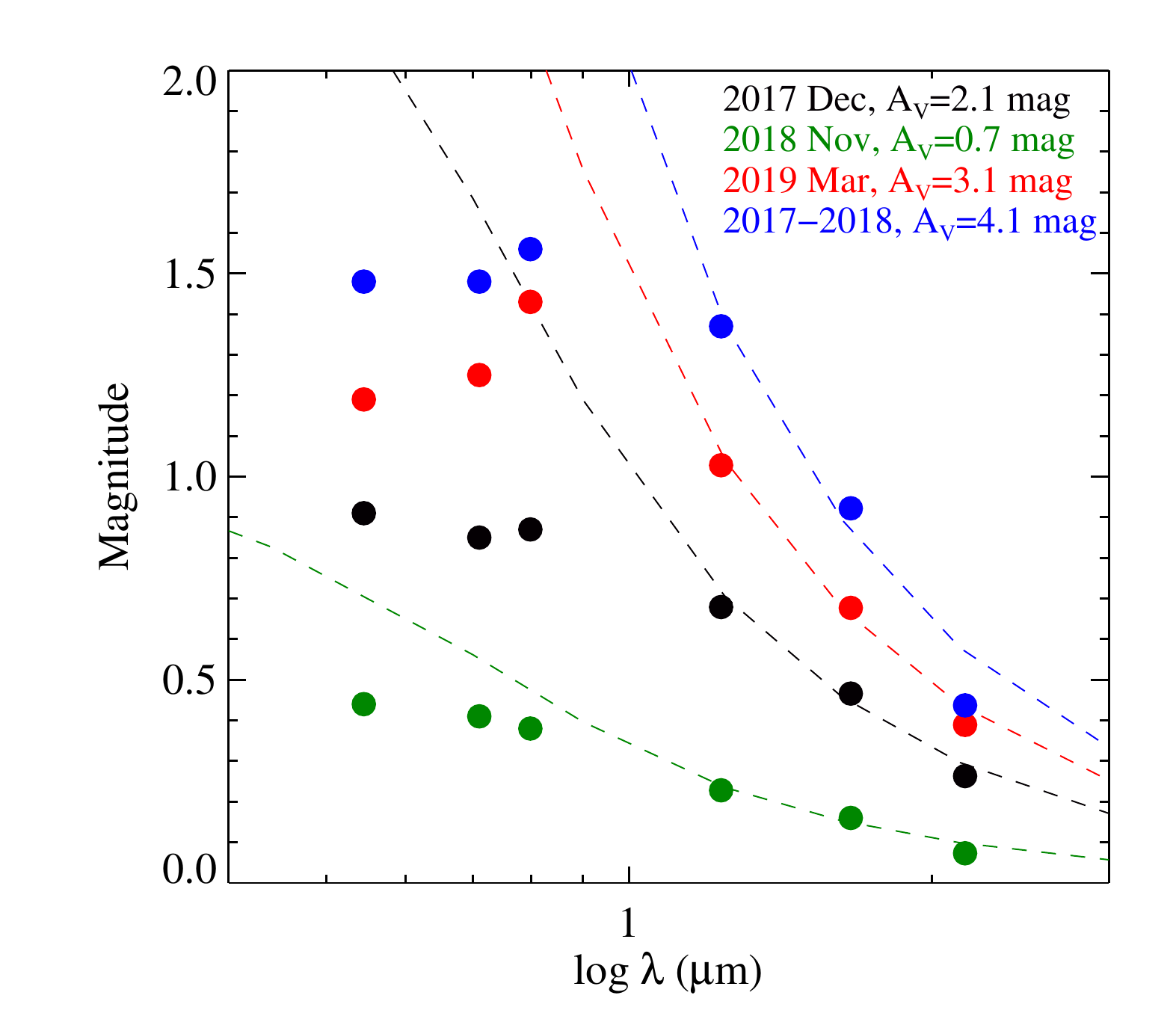}
 \caption{
 Magnitude differences between the minimum and maximum of the $\sim$5 day brightness variations in 2017 Dec (black), in 2018 Nov (green), and in 2019 Mar (red). The blue symbols show the difference between the brightness during outburst (2017 Dec) and quiescence (2018 Nov), and both correspond to the same phase (maximum) of the 5-day brightness variations. The visual extinctions corresponding to the $J$, $H$, and $K_{\rm{S}}$ data and the extinction curves calculated using the formulae from \citet{cardelli} are also shown.
 }
 \label{mag_difference}
\end{figure}

\begin{figure}
 \includegraphics[height=\columnwidth,angle=90]{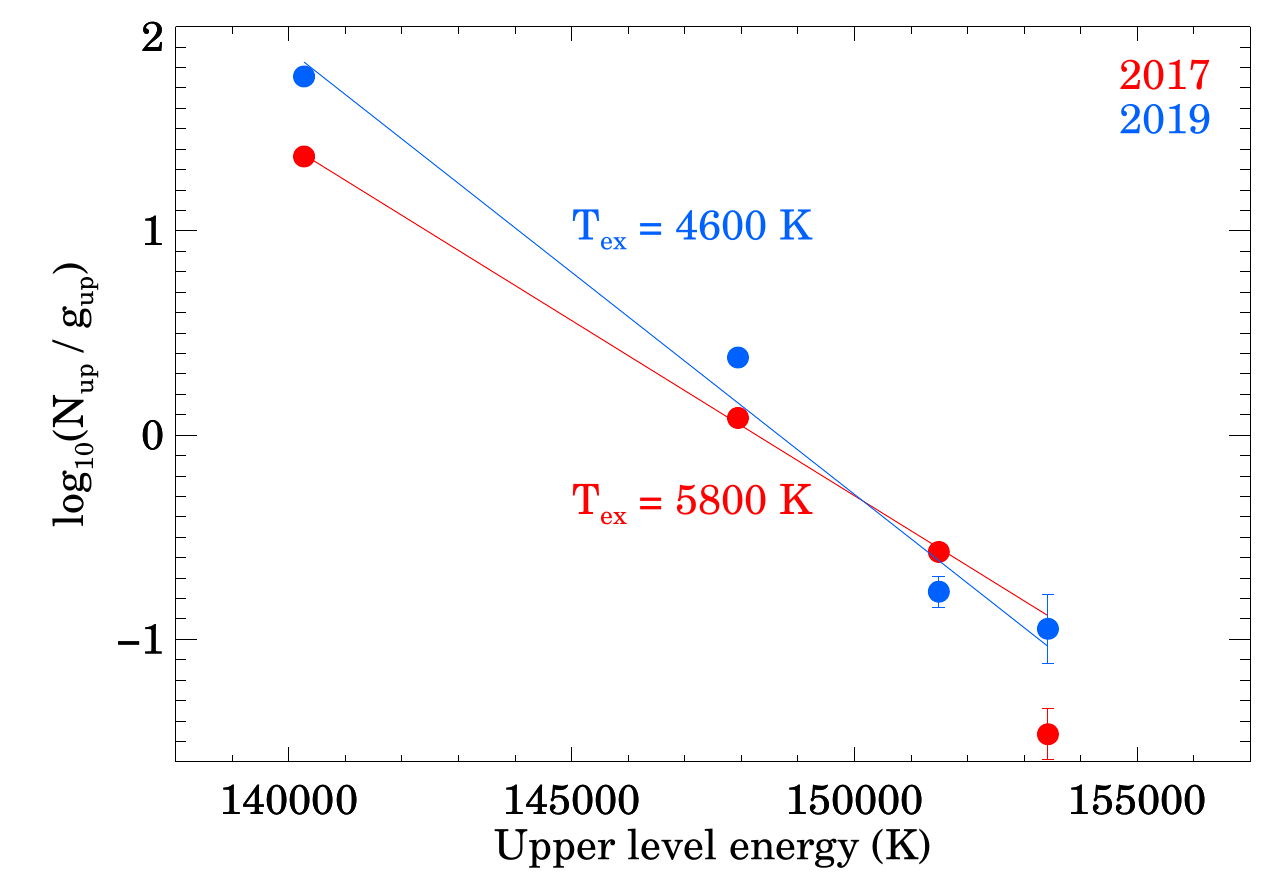}
 \caption{Excitation diagram for hydrogen. The linear fits only use the H$\alpha$, H$\beta$, and H$\gamma$ lines.}
 \label{fig:hydrogen}
\end{figure}

In addition to extinction, mass accretion from the circumstellar disc onto the star may also cause brightness variations. Based on our optical spectra obtained close in time to the brightest and the faintest epochs, we derived accretion rates from the flux of the H$\alpha$ line, taking its broad components as accretion tracer. The measured line flux values are presented in Tab.~\ref{tab:lines}, but for our analysis they need to be corrected for extinction. Based on our previous result, for the 2017 December 17 spectrum (bright state) we adopted A$_V$=2.3~mag and used the standard extinction law of \citet{cardelli}. This calculation resulted in 4.2$\times$10$^{-13}$ erg cm$^{-2}$ s$^{-1}$. Correcting the line fluxes in the faint state is less straightforward, since at these wavelengths the measured variability amplitude is lower than predicted by the extinction law, as seen in Fig. \ref{mag_difference}. 
Thus, using the A$_V$=6.4~mag and the \citet{cardelli} extinction law would over-predict the extinction corrected line flux of the H$\alpha$ line. 
Instead, we will assume that the H$\alpha$ line flux and the neighbouring continuum changed, due to extinction, by the same amount between the bright and faint epochs. 
Taking advantage that our spectra are flux calibrated (Sect.~\ref{sec:spectroscopy}), we determined a ratio of 3.0 between the continuum fluxes in the bright and faint spectra around 656~nm, and scaled up the H$\alpha$ line flux measured in 2019 by this ratio. It yielded 1.0$\times$10$^{-12}$ erg cm$^{-2}$ s$^{-1}$. Then we used the equations of \citet{Alcala2014} to transform L$_{\rm line}$ to L$_{\rm acc}$, and finally to \.M$_{\rm acc}$ (adopting M$_{\rm star}=1.1$ M$_{\odot}$, and  R$_{\rm star}$=3R$_{\odot}$). The results are \.M$_{\rm acc}$ = 3.3$\times$10$^{-9}$~M$_{\odot}$yr$^{-1}$ and 8.8$\times$10$^{-9}$~M$_{\odot}$yr$^{-1}$ in 2017 December and in 2019 March, respectively.

Detecting several lines of the Balmer series (Sect.~\ref{sec:spectroscopy}) allows us to determine the excitation temperature of the gas that emits these lines. First, we corrected the line fluxes in both the bright and faint states, following the procedure applied for the H$\alpha$ line above. 
Then in Fig.~\ref{fig:hydrogen} we plotted the relative level populations calculated in LTE approximation divided by the statistical weights of the energy levels in question in logarithmic scale as a function of the energy of the levels. The data point for H$\delta$ is quite uncertain, so we fitted only three data points, H$\alpha$, H$\beta$, and H$\gamma$, with a straight line. The inverse of the slope of this line gave an excitation temperature of 5800\,K in 2017 and 4600\,K in 2019. These values are higher than the photospheric temperature of a K4-type star, indicating that the detected hydrogen lines originate from gas participating in the accretion process, located close to the magnetospheric hot spot, or residing in the accretion column.
The excitation temperatures calculated above are similar to those calculated by \citet{Muzerolle1998b} in their magnetospheric accretion models.
We also used Eqn. 9 from \citet{Hartmann2016} to calculate the effective temperature of the heated photosphere of V555 Ori. Assuming the average value of the derived accretion rates, and the result is $\sim$5500 K.

The calculated accretion rate change would imply a change in the optical veiling, as these two quantities are expected to correlate well \citep[e.g.,][]{muzerolle1998}. However, the equivalent width of the Li absorption line was constant within the measurement uncertainties, suggesting constant veiling. This apparent contradiction can be resolved if we consider that the accretion luminosities calculated for 2017 Dec and 2019 Mar (L$_{\rm acc}$=0.03 L$_{\odot}$ and 0.08 L$_{\odot}$, respectively) are so low compared to the luminosity of the star, that the resulting veiling (and its change) would remain below our detection limit (cf. Fig.~4 of \citealt{muzerolle1998}). Also, the luminosity changes caused by accretion variability would not be enough to explain the observed brightness variations of V555~Ori, leaving us with the most plausible interpretation that the light curves in Fig. \ref{lightcurve_gaia_wise} are related to changing obscuration along the line of sight rather than to accretion changes.

\subsection{The circumstellar structure of V555~Ori}

\begin{figure}
 \includegraphics[width=\columnwidth]{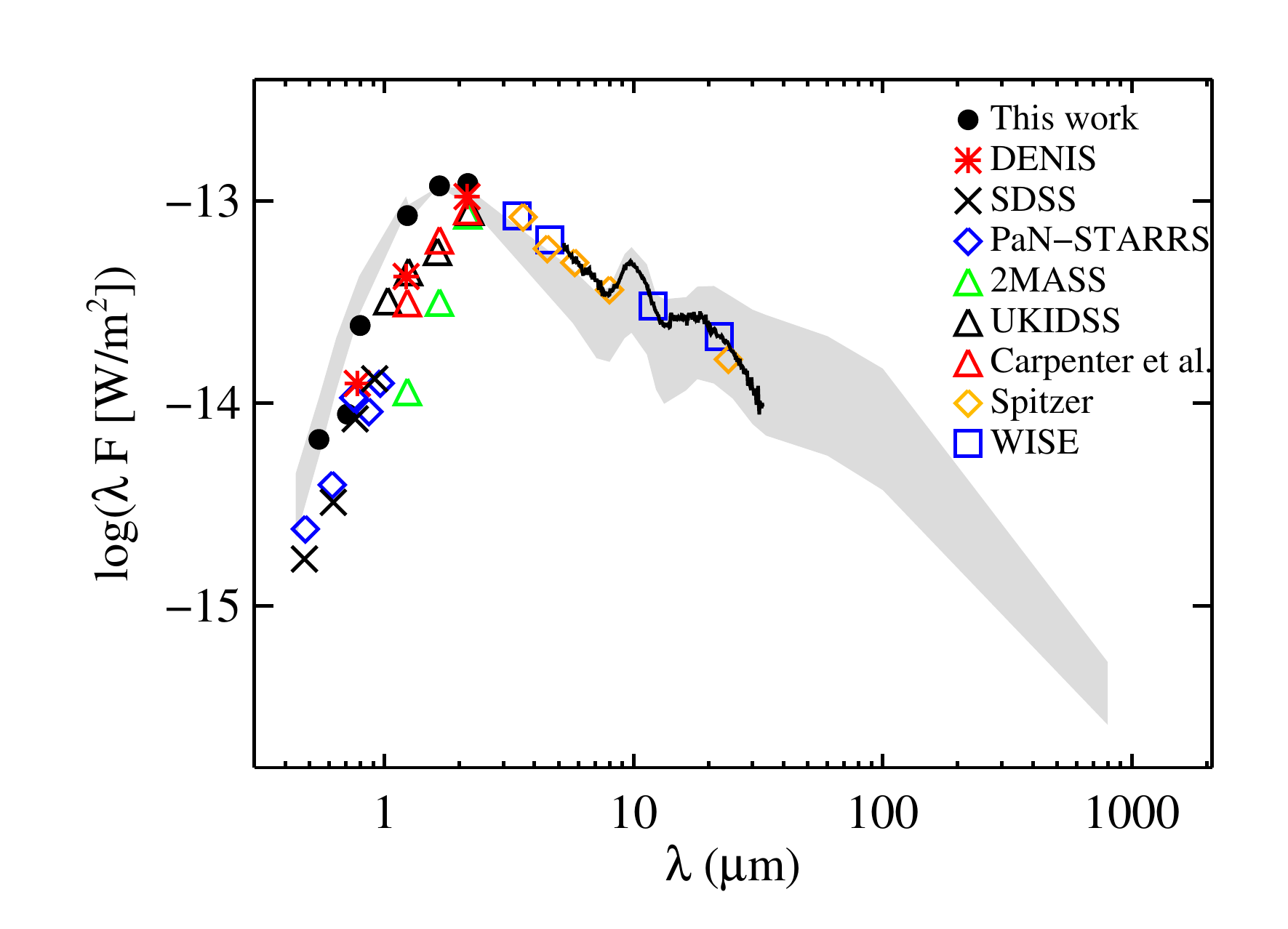}
 \caption{Spectral energy distribution of V555 Ori. The shaded area shows the `Taurus' median SED reddened by the extinction of $A_V$=2.3 mag determined for the bright state and normalised to the H-band flux measured with SMARTS in the bright state (black circles).
The data points below 1.25 $\mu$m and above 34 $\mu$m are from \citet{DAlessio1999}; otherwise from \citet{Furlan2006}.
The spectrum, obtained with the Infrared Spectrograph (IRS) of \textit{Spitzer} was downloaded from the Combined Atlas of Sources with \textit{Spitzer} IRS Spectra.
}
 \label{sed}
\end{figure}

Based on the available photometric data and the reddening values determined in the previous subsection, we constructed the spectral energy distribution (SED) of V555 Ori over the 0.36$-$24~$\micron$ wavelength range, as seen in Fig. \ref{sed}. In addition to data collected from the literature, we also plotted the optical--infrared fluxes obtained by us at the date of the maximum brightness on 2017 December 19. The large scatter in the optical and near-infrared parts is consistent with the large-amplitude variability analysed in this work. While the optical part is dominated by the reddened star, the shape of the SED at infrared wavelengths suggests the presence of a circumstellar disc. The \textit{Spitzer} IRS spectrum also reveals broad 10~$\mu$m and 20~$\mu$m spectral features attributed to small silicate grains in the surface layer of the disc. In the figure we overplotted the ``Taurus median'', which is based on the SEDs of T Tauri stars in the Taurus region (\citealp{DAlessio1999}, \citealp{Furlan2006}). We reddened the Taurus median by $A_V$=2.3 mag, as derived above for the bright state, and normalized it to the H-band flux of V555~Ori. The figure shows that the Taurus median is very similar to the measured SED, indicating that V555~Ori possesses a rather typical, average protoplanetary disc. While such discs are usually passive, i.e. powered mainly by irradiation from the central T~Tauri star, a smaller contribution from the accretion process may also present.

\subsection{The origin of the short-term variability}
\label{sec:shortterm}

Based on observations carried out with SMARTS and TESS, we identified brightness variations with an amplitude of $\sim$1 mag in V band on a time-scale of $\sim$5 days. In order to determine the physical origin of this variability, in Fig.~\ref{mag_difference} we compared the multicolour magnitude differences corresponding to the short-term ($\sim$5-day) and long term brightness variations, based on our SMARTS data in the $V$, $R_{C}$, $I_{C}$, $J$, $H$, and K$_S$ bands. The short-term brightness differences represent the magnitude change between the maximum and minimum of a selected $\sim$5-day cycle. The plotted long-term variation corresponds to the difference between the brightest state (2017 December 19) and the faintest state (2018 November 19), and both correspond to the maximum phase of the 5-day brightness variations. The figure clearly demonstrates that the wavelength dependence of the short-term and long-term variations are indistinguishable. They differ only in their amplitudes, the short-term amplitudes being somewhat smaller. In all cases, the $J$, $H$, and $K_S$ magnitude differences closely follow the standard extinction curve (overplotted in the figure). The optical magnitude differences, however, always smaller than predicted from the extinction curve matching the near-IR data points of the same date. Moreover, the optical changes show little wavelength dependence, as was already implied by the left panel of Fig.~5. One possible reason of the lower optical magnitude differences could be a significant scattered light contribution that originates from the outer disc regions. The presence of an invariable optical component would render any changes related to obscuration of a part of the system smaller on the magnitude scale. Our result strongly suggests that the brightness variations on both short and long time-scales are related to changing extinction. 
Given that the accretion luminosities are low (0.03-0.08 $L_\odot$) compared to the luminosity of V555 Ori (3.39 $L_\odot$), the short-term brightness variations of 1 mag - equivalent to a factor of 2.5 change in the luminosity - are unlikely to be affected by the accretion.

\subsection{A twin of AA Tau?}

Quasi-periodic variability, caused by changing extinction, is a defining characteristics of the dipper class of variable low-mass young stellar objects. Brightness variations, similar to those observed in V555~Ori, were found earlier for AA~Tau, the prototypical member of the dipper class: \citet{Bouvier1999} observed a $\sim$8.5 day, $\sim$1.4 mag quasi-periodic variability in the $U$,$B$,$V$,$R$, and $I$ filters. The behavior of AA~Tau was interpreted as periodic occultations of the star by an inner disc warp, and are likely driven by a misalignment between the stellar magnetic dipole and rotation axes \citep{Bouvier1999}. By now a large sample of AA Tau-type stars were identified, e.g. \citet{Rice2015} discovered 73 periodic AA Tau type variables in the Orion Nebula Cluster, based on $J$, $H$, and $K$ colours. In a sample of 83 T Tau stars, \citet{Alencar2010} found that at least $\sim$28\% of them show light variations similar to AA~Tau.

The extinction-related brightness variations suggest that V555 Ori is an AA Tau-like star, and its short-term periodic variability can be explained in terms of the dipper model. While this scenario would not account for the long-term changes, we recall that in 2011 AA~Tau underwent a sudden dimming, which was likely a result of a density perturbation in the disc at a somewhat larger radius than the inner disc warp \citep{Bouvier2013}. The optical colours of AA Tau were wavelength-independent, also during the dimming (\citealp{Bouvier1999}, \citealp{Bouvier2003}), similar to the case of V555 Ori (e.g. Fig. \ref{mag_difference}), which can be explained as the emission at these wavelengths is dominated by scattered light.

Similarities between the optical spectra of V555 Ori and AA Tau further support that V555~Ori is an AA Tau-type star. Based on \citet{Bouvier2013}, several lines were detected toward both stars in emission, such as the hydrogen Balmer lines from H$\alpha$ to H$\delta$, the [OI] 630.0/636.4 nm lines, the [NII] 658.3 nm line, and the [SII] 673.1 nm line. Some additional lines were detected toward AA Tau, but not toward V555 Ori, such as HeI and CaII lines.

While V555 Ori shows strong evidences to be an AA Tau-like star, there are also a few interesting differences. In AA~Tau, no  quasi-periodic brightness variations can be seen in the long-term deep state, although they were present for 20 years before \citep{Bouvier2013}. Such periodic changes, however, are continuously apparent in our observations of V555 Ori, even in the faint state. 

The other difference is that in V555~Ori the eclipses, both the periodic ones caused by the warp and the long-term ones caused by a more distant density enhancement in the disc, are associated with surprisingly invariable optical colours. It may suggest that the obscuring material is more optically thick than in AA Tau, and also that the transition from optically thin and optically thick regions (the edge of the obscuring structure) is very sharp.

If the interpretation of the variability of V555 Ori by the dipper model is correct, then part of the disc periodically ($P \sim$5 days) intercepts the line of sight to the star, and causes the fading episodes. Assuming Keplerian rotation, we can estimate the radial distance ($r_c$) of the clump which causes the 5-day brightness variations using
\begin{equation}
\nonumber
r_c = (P / 2 \pi)^{2/3} (G M_\star)^{1/3} = 12.7 R_\odot = 3.9 R_\star    
\end{equation}
with $M_\star \sim $1.1 $M_\odot$ \citep{DaRio2016} and $R_\star \sim 3.3 R_\odot$ (based on its effective temperature of $\sim$4300~K and its luminosity of $L_\star=3.39 L_\odot$ reported by \citealp{DaRio2016}). Similar to \citet{Bouvier1999} we can calculate the dust temperature at the derived distance from the star as
\begin{equation}
\nonumber
T_d=\left( \frac{L_\star}{16 \pi \sigma r_c^2} \right)^{1/4} \sim 1500~{\rm{K}}.
\end{equation}
This is consistent with the sublimation temperature of $T_S\sim1550$ K at the derived $r_c$ (e.g. \citealp{MonnierMillanGabet2002}).
The duration of the dips in the light curve is approximately half the photometric period of 5 days. This is similar to the case of AA Tau \citep{Bouvier1999}, and implies, that the obscuring material is elongated azimuthally, as it extends over $\pi$ radians around the central star.

\begin{figure}
 \includegraphics[width=\columnwidth]{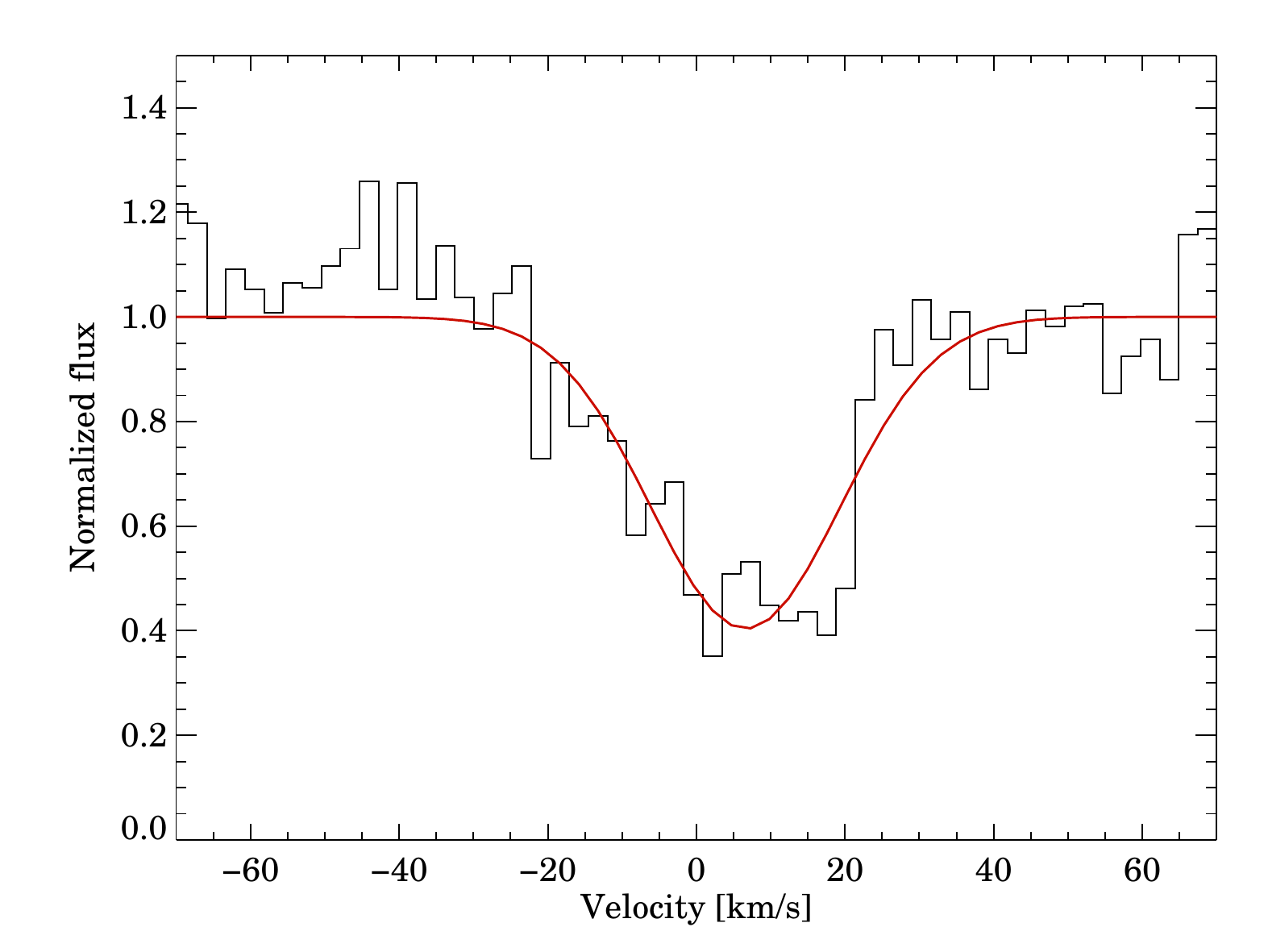}
 \caption{The normalized average \ion{Li}{i} line profile with a Gaussian fit.}
 \label{fig:Li}
\end{figure}

Dippers are generally seen at high inclination, close to edge-on orientation \citep{Alencar2010}, including AA Tau with an inclination of 75$^\circ$ \citep{Bouvier1999}. There are indirect tracers which may suggest a high inclination for V555 Ori as well.
In T Tau stars, the [\ion{O}{i}] line has a low-velocity component, and a strongly blue-shifted component which originates in a high velocity jet. The lack of such a high velocity [\ion{O}{i}] component toward AA Tau led to the conclusion that it is due to a close to edge-on orientation, where the jet lies in the plane of the sky \citep{Bouvier1999}. The lack of a high velocity [\ion{O}{i}] component is also seen for V555 Ori, and might be a result of a close to edge-on orientation.
A more direct approach to estimate the inclination is using a measurement of vsin$i$. 
The vsin$i$ measured for V555 Ori by \citet{DaRio2016} is 26.70$\pm$22.01 km s$^{-1}$. Assuming a rotational period of 5 days, the vsin$i$ value of 26.70 km s$^{-1}$ results in an inclination of $\sim$53$^\circ$. We can also use our FEROS measurement of the \ion{Li}{i} absorption line to measure the vsin$i$ (Fig. \ref{fig:Li}). To convert the measured FWHM line width of the average \ion{Li}{i} line (30.1$\pm$2.5 km/s) to vsin$i$, we compared it to FWHM line widths of the same line toward other young stars also measured with FEROS, which have already derived vsin$i$ values \citep{Weise2010}. Based on this comparison, the vsin$i$ was found to be $18.2^{+8.7}_{-5.6}$ km s$^{-1}$, where the best fit value corresponds to the average \ion{Li}{i} line profile, and the error bars were derived based on the range defined by fitting the measured individual spectra. Assuming a rotational period of 5 days, these values suggest an inclination of $33.4^{+20.3}_{-11.2}$ degrees, therefore V555 Ori might be a dipper seen at a lower inclination.

The formation of warps in the discs of T Tau stars was also confirmed by magnetohydrodynamic (MHD) simulations. The MHD simulations of \citet{Romanova2013} were able to reproduce the warp properties suggested for AA Tau with a simulation run that forms a large amplitude warp that rotates with the frequency of the star in their model FW$\mu$1.5 (for the details of this model see \citealp{Romanova2013}). Fig. 4 of \citet{Romanova2013} shows warps at various radial distances from the star, including one at $\sim$4.3 $R_\star$, which is close to the value of 3.9 $R_\star$ derived above for V555 Ori.

\section{Summary}
\label{sect:summary}

We analysed photometric and spectroscopic data of the {\it Gaia} alerted T Tau star V555 Ori. The main results can be summarized as follows.

\begin{itemize}

\item[-] We found both long-term (on a time-scale of months / years) and short-term quasi-periodic (with a period of $\sim$5 days) brightness variations. 
At optical wavelengths both the long-term and short-term variations exhibited colourless changes over the whole measured brightness range, while in the near-infrared they were consistent with changing extinction. This behaviour makes V555 Ori similar to the pre-main sequence star AA Tau, the prototype of the 'dipper' variable class, where the light changes are interpreted as periodic eclipses of the star by a rotating inner disc warp.

\item[-] Unlikely in AA Tau, the 5 days periodic behaviour was also continuously detectable in the faint phase,
implying that the inner disc warp remained stable in both the high and low states of the system.

\item[-] The long-term variability was probably also due to
extinction. At the brightness maximum seen in 2017, V555 Ori was in a moderately obscured ($A_V=2.3$ mag) state, while in the low state the extinction in the V band was as high as 6.4 mag.

\item[-] Several emission lines were detected in the optical spectra (measured in 2017 and 2019) of V555 Ori: lines of the hydrogen Balmer series from H$\alpha$ to H$\delta$, two forbidden ionized sulphur lines, two forbidden ionized nitrogen lines, and two forbidden neutral oxygen lines.
The changes in the accretion rate derived from the H$\alpha$ line revealed temporary changes, but its magnitude is insufficient to explain the observed optical flux changes.

\item[-] Based on archive data and the data obtained with SMARTS, the SED of V555 Ori is typical of other T Tau stars, based on a comparison to the `Taurus median' SED.

\end{itemize}

\section*{Acknowledgements}

We thank the referee for comments which helped to improve our manuscript.
This project has received funding from the European Research Council (ERC) under the European Union's Horizon 2020 research and innovation programme under grant agreement No 716155 (SACCRED).
The project has also received funding  from the ERC under grant agreement No 742095 (SPIDI).
This project has been supported by the Lend\"ulet Program  of the Hungarian Academy of Sciences, project No. LP2018-7.
ESA PRODEX contract nr. 4000129910.
This project has been supported by the GINOP-2.3.2-15-2016-00003 grant of the Hungarian National Research, Development and Innovation Office (NKFIH).
ZsMSz is supported by the \'UNKP-20-2 New National Excellence Program of the Ministry for Innovation and Technology from the source of the National Research, Development and Innovation Fund.
LK is supported by the Bolyai J\'anos Research Scholarship of the Hungarian Academy of Sciences.
The authors acknowledge the Hungarian National Research, Development and Innovation Office grants OTKA KH-130526, K-131508, PD-134784, and 119993.
This paper includes data collected by the TESS mission. Funding for the TESS
mission is provided by the NASA Explorer Program.
We thank John Carpenter for $JHK$ data of V555 Ori.

\section*{Data availability}

The data underlying this article will be shared on reasonable request to the corresponding author.




\bibliographystyle{mnras}
\bibliography{v555ori} 

\begin{thebibliography}{}
\makeatletter
\relax
\def\mn@urlcharsother{\let\do\@makeother \do\$\do\&\do\#\do\^\do\_\do\%\do\~}
\def\mn@doi{\begingroup\mn@urlcharsother \@ifnextchar [ {\mn@doi@}
  {\mn@doi@[]}}
\def\mn@doi@[#1]#2{\def\@tempa{#1}\ifx\@tempa\@empty \href
  {http://dx.doi.org/#2} {doi:#2}\else \href {http://dx.doi.org/#2} {#1}\fi
  \endgroup}
\def\mn@eprint#1#2{\mn@eprint@#1:#2::\@nil}
\def\mn@eprint@arXiv#1{\href {http://arxiv.org/abs/#1} {{\tt arXiv:#1}}}
\def\mn@eprint@dblp#1{\href {http://dblp.uni-trier.de/rec/bibtex/#1.xml}
  {dblp:#1}}
\def\mn@eprint@#1:#2:#3:#4\@nil{\def\@tempa {#1}\def\@tempb {#2}\def\@tempc
  {#3}\ifx \@tempc \@empty \let \@tempc \@tempb \let \@tempb \@tempa \fi \ifx
  \@tempb \@empty \def\@tempb {arXiv}\fi \@ifundefined
  {mn@eprint@\@tempb}{\@tempb:\@tempc}{\expandafter \expandafter \csname
  mn@eprint@\@tempb\endcsname \expandafter{\@tempc}}}

\bibitem[\protect\citeauthoryear{{Alcal{\'a}} et~al.,}{{Alcal{\'a}}
  et~al.}{2014}]{Alcala2014}
{Alcal{\'a}} J.~M.,  et~al., 2014, \mn@doi [\aap]
  {10.1051/0004-6361/201322254}, \href
  {https://ui.adsabs.harvard.edu/abs/2014A&A...561A...2A} {561, A2}

\bibitem[\protect\citeauthoryear{{Alencar} et~al.,}{{Alencar}
  et~al.}{2010}]{Alencar2010}
{Alencar} S.~H.~P.,  et~al., 2010, \mn@doi [\aap]
  {10.1051/0004-6361/201014184}, \href
  {https://ui.adsabs.harvard.edu/abs/2010A&A...519A..88A} {519, A88}

\bibitem[\protect\citeauthoryear{{Bailer-Jones}, {Rybizki}, {Fouesneau},
  {Mantelet}  \& {Andrae}}{{Bailer-Jones} et~al.}{2018}]{BailerJones2018}
{Bailer-Jones} C.~A.~L.,  {Rybizki} J.,  {Fouesneau} M.,  {Mantelet} G.,
  {Andrae} R.,  2018, \mn@doi [\aj] {10.3847/1538-3881/aacb21}, \href
  {https://ui.adsabs.harvard.edu/abs/2018AJ....156...58B} {156, 58}

\bibitem[\protect\citeauthoryear{{Bessell} \& {Brett}}{{Bessell} \&
  {Brett}}{1988}]{loci}
{Bessell} M.~S.,  {Brett} J.~M.,  1988, \mn@doi [\pasp] {10.1086/132281}, \href
  {http://adsabs.harvard.edu/abs/1988PASP..100.1134B} {100, 1134}

\bibitem[\protect\citeauthoryear{{Bouvier} et~al.,}{{Bouvier}
  et~al.}{1999}]{Bouvier1999}
{Bouvier} J.,  et~al., 1999, \aap, \href
  {https://ui.adsabs.harvard.edu/abs/1999A&A...349..619B} {349, 619}

\bibitem[\protect\citeauthoryear{{Bouvier} et~al.,}{{Bouvier}
  et~al.}{2003}]{Bouvier2003}
{Bouvier} J.,  et~al., 2003, \mn@doi [\aap] {10.1051/0004-6361:20030938}, \href
  {https://ui.adsabs.harvard.edu/abs/2003A&A...409..169B} {409, 169}

\bibitem[\protect\citeauthoryear{{Bouvier}, {Grankin}, {Ellerbroek}, {Bouy}  \&
  {Barrado}}{{Bouvier} et~al.}{2013}]{Bouvier2013}
{Bouvier} J.,  {Grankin} K.,  {Ellerbroek} L.~E.,  {Bouy} H.,   {Barrado} D.,
  2013, \mn@doi [\aap] {10.1051/0004-6361/201321389}, \href
  {https://ui.adsabs.harvard.edu/abs/2013A&A...557A..77B} {557, A77}

\bibitem[\protect\citeauthoryear{{Brahm}, {Jord{\'a}n}  \& {Espinoza}}{{Brahm}
  et~al.}{2017}]{Brahm2017}
{Brahm} R.,  {Jord{\'a}n} A.,   {Espinoza} N.,  2017, \mn@doi [\pasp]
  {10.1088/1538-3873/aa5455}, \href
  {https://ui.adsabs.harvard.edu/abs/2017PASP..129c4002B} {129, 034002}

\bibitem[\protect\citeauthoryear{{Canto}, {Meaburn}, {Theokas}  \&
  {Elliott}}{{Canto} et~al.}{1980}]{Canto1980}
{Canto} J.,  {Meaburn} J.,  {Theokas} A.~C.,   {Elliott} K.~H.,  1980, \mn@doi
  [\mnras] {10.1093/mnras/193.4.911}, \href
  {https://ui.adsabs.harvard.edu/abs/1980MNRAS.193..911C} {193, 911}

\bibitem[\protect\citeauthoryear{{Cardelli}, {Clayton}  \& {Mathis}}{{Cardelli}
  et~al.}{1989}]{cardelli}
{Cardelli} J.~A.,  {Clayton} G.~C.,   {Mathis} J.~S.,  1989, \mn@doi [\apj]
  {10.1086/167900}, \href {http://adsabs.harvard.edu/abs/1989ApJ...345..245C}
  {345, 245}

\bibitem[\protect\citeauthoryear{{Carpenter}, {Hillenbrand}  \&
  {Skrutskie}}{{Carpenter} et~al.}{2001}]{carpenter}
{Carpenter} J.~M.,  {Hillenbrand} L.~A.,   {Skrutskie} M.~F.,  2001, \mn@doi
  [\aj] {10.1086/321086}, \href
  {http://adsabs.harvard.edu/abs/2001AJ....121.3160C} {121, 3160}

\bibitem[\protect\citeauthoryear{{D'Alessio}, {Calvet}, {Hartmann}, {Lizano}
  \& {Cant{\'o}}}{{D'Alessio} et~al.}{1999}]{DAlessio1999}
{D'Alessio} P.,  {Calvet} N.,  {Hartmann} L.,  {Lizano} S.,   {Cant{\'o}} J.,
  1999, \mn@doi [\apj] {10.1086/308103}, \href
  {https://ui.adsabs.harvard.edu/abs/1999ApJ...527..893D} {527, 893}

\bibitem[\protect\citeauthoryear{{Da Rio} et~al.,}{{Da Rio}
  et~al.}{2016}]{DaRio2016}
{Da Rio} N.,  et~al., 2016, \mn@doi [\apj] {10.3847/0004-637X/818/1/59}, \href
  {https://ui.adsabs.harvard.edu/abs/2016ApJ...818...59D} {818, 59}

\bibitem[\protect\citeauthoryear{{Dere}, {Landi}, {Mason}, {Monsignori Fossi}
  \& {Young}}{{Dere} et~al.}{1997}]{dere1997}
{Dere} K.~P.,  {Landi} E.,  {Mason} H.~E.,  {Monsignori Fossi} B.~C.,   {Young}
  P.~R.,  1997, \mn@doi [\aaps] {10.1051/aas:1997368}, \href
  {https://ui.adsabs.harvard.edu/abs/1997A&AS..125..149D} {125, 149}

\bibitem[\protect\citeauthoryear{{Dere}, {Del Zanna}, {Young}, {Landi}  \&
  {Sutherland}}{{Dere} et~al.}{2019}]{dere2019}
{Dere} K.~P.,  {Del Zanna} G.,  {Young} P.~R.,  {Landi} E.,   {Sutherland}
  R.~S.,  2019, \mn@doi [\apjs] {10.3847/1538-4365/ab05cf}, \href
  {https://ui.adsabs.harvard.edu/abs/2019ApJS..241...22D} {241, 22}

\bibitem[\protect\citeauthoryear{{Dullemond}, {van den Ancker}, {Acke}  \& {van
  Boekel}}{{Dullemond} et~al.}{2003}]{Dullemond2003}
{Dullemond} C.~P.,  {van den Ancker} M.~E.,  {Acke} B.,   {van Boekel} R.,
  2003, \mn@doi [\apjl] {10.1086/378400}, \href
  {https://ui.adsabs.harvard.edu/abs/2003ApJ...594L..47D} {594, L47}

\bibitem[\protect\citeauthoryear{{Furlan} et~al.,}{{Furlan}
  et~al.}{2006}]{Furlan2006}
{Furlan} E.,  et~al., 2006, \mn@doi [\apjs] {10.1086/505468}, \href
  {https://ui.adsabs.harvard.edu/abs/2006ApJS..165..568F} {165, 568}

\bibitem[\protect\citeauthoryear{{Gomez de Castro} \& {Pudritz}}{{Gomez de
  Castro} \& {Pudritz}}{1993}]{GomezdeCastroPudritz1993}
{Gomez de Castro} A.~I.,  {Pudritz} R.~E.,  1993, \mn@doi [\apj]
  {10.1086/172704}, \href
  {https://ui.adsabs.harvard.edu/abs/1993ApJ...409..748G} {409, 748}

\bibitem[\protect\citeauthoryear{{Hartmann} \& {Kenyon}}{{Hartmann} \&
  {Kenyon}}{1996}]{HartmannKenyon1996}
{Hartmann} L.,  {Kenyon} S.~J.,  1996, \mn@doi [\araa]
  {10.1146/annurev.astro.34.1.207}, \href
  {https://ui.adsabs.harvard.edu/abs/1996ARA&A..34..207H} {34, 207}

\bibitem[\protect\citeauthoryear{{Hartmann}, {Herczeg}  \& {Calvet}}{{Hartmann}
  et~al.}{2016}]{Hartmann2016}
{Hartmann} L.,  {Herczeg} G.,   {Calvet} N.,  2016, \mn@doi [\araa]
  {10.1146/annurev-astro-081915-023347}, \href
  {https://ui.adsabs.harvard.edu/abs/2016ARA&A..54..135H} {54, 135}

\bibitem[\protect\citeauthoryear{{Herbig}}{{Herbig}}{1977}]{Herbig1977}
{Herbig} G.~H.,  1977, \mn@doi [\apj] {10.1086/155615}, \href
  {https://ui.adsabs.harvard.edu/abs/1977ApJ...217..693H} {217, 693}

\bibitem[\protect\citeauthoryear{{Hollenbach}}{{Hollenbach}}{1997}]{Hollenbach1997}
{Hollenbach} D.,  1997, in {Reipurth} B.,  {Bertout} C.,  eds, ~ Vol. 182,
  Herbig-Haro Flows and the Birth of Stars. pp 181--198

\bibitem[\protect\citeauthoryear{{Hollenbach} \& {Tielens}}{{Hollenbach} \&
  {Tielens}}{1997}]{HollenbachTielens1997}
{Hollenbach} D.~J.,  {Tielens} A.~G.~G.~M.,  1997, \mn@doi [\araa]
  {10.1146/annurev.astro.35.1.179}, \href
  {https://ui.adsabs.harvard.edu/abs/1997ARA&A..35..179H} {35, 179}

\bibitem[\protect\citeauthoryear{{Kaufer}, {Stahl}, {Tubbesing},
  {N{\o}rregaard}, {Avila}, {Francois}, {Pasquini}  \& {Pizzella}}{{Kaufer}
  et~al.}{1999}]{kaufer}
{Kaufer} A.,  {Stahl} O.,  {Tubbesing} S.,  {N{\o}rregaard} P.,  {Avila} G.,
  {Francois} P.,  {Pasquini} L.,   {Pizzella} A.,  1999, The Messenger, \href
  {http://adsabs.harvard.edu/abs/1999Msngr..95....8K} {95, 8}

\bibitem[\protect\citeauthoryear{{Meyer}, {Calvet}  \& {Hillenbrand}}{{Meyer}
  et~al.}{1997}]{meyer1997}
{Meyer} M.~R.,  {Calvet} N.,   {Hillenbrand} L.~A.,  1997, \mn@doi [\aj]
  {10.1086/118474}, \href {http://adsabs.harvard.edu/abs/1997AJ....114..288M}
  {114, 288}

\bibitem[\protect\citeauthoryear{{Monnier} \& {Millan-Gabet}}{{Monnier} \&
  {Millan-Gabet}}{2002}]{MonnierMillanGabet2002}
{Monnier} J.~D.,  {Millan-Gabet} R.,  2002, \mn@doi [\apj] {10.1086/342917},
  \href {https://ui.adsabs.harvard.edu/abs/2002ApJ...579..694M} {579, 694}

\bibitem[\protect\citeauthoryear{{Morales-Calder{\'o}n}
  et~al.,}{{Morales-Calder{\'o}n} et~al.}{2011}]{MoralesCalderon2011}
{Morales-Calder{\'o}n} M.,  et~al., 2011, \mn@doi [\apj]
  {10.1088/0004-637X/733/1/50}, \href
  {https://ui.adsabs.harvard.edu/abs/2011ApJ...733...50M} {733, 50}

\bibitem[\protect\citeauthoryear{{Muzerolle}, {Hartmann}  \&
  {Calvet}}{{Muzerolle} et~al.}{1998a}]{muzerolle1998}
{Muzerolle} J.,  {Hartmann} L.,   {Calvet} N.,  1998a, \mn@doi [\aj]
  {10.1086/300428}, \href
  {https://ui.adsabs.harvard.edu/abs/1998AJ....116..455M} {116, 455}

\bibitem[\protect\citeauthoryear{{Muzerolle}, {Calvet}  \&
  {Hartmann}}{{Muzerolle} et~al.}{1998b}]{Muzerolle1998b}
{Muzerolle} J.,  {Calvet} N.,   {Hartmann} L.,  1998b, \mn@doi [\apj]
  {10.1086/305069}, \href
  {https://ui.adsabs.harvard.edu/abs/1998ApJ...492..743M} {492, 743}

\bibitem[\protect\citeauthoryear{{P{\'a}l}}{{P{\'a}l}}{2012}]{Pal2012}
{P{\'a}l} A.,  2012, \mn@doi [\mnras] {10.1111/j.1365-2966.2011.19813.x}, \href
  {https://ui.adsabs.harvard.edu/abs/2012MNRAS.421.1825P} {421, 1825}

\bibitem[\protect\citeauthoryear{{Rice}, {Reipurth}, {Wolk}, {Vaz}  \&
  {Cross}}{{Rice} et~al.}{2015}]{Rice2015}
{Rice} T.~S.,  {Reipurth} B.,  {Wolk} S.~J.,  {Vaz} L.~P.,   {Cross} N.~J.~G.,
  2015, \mn@doi [\aj] {10.1088/0004-6256/150/4/132}, \href
  {https://ui.adsabs.harvard.edu/abs/2015AJ....150..132R} {150, 132}

\bibitem[\protect\citeauthoryear{{Ricker} et~al.,}{{Ricker}
  et~al.}{2015}]{Ricker2015}
{Ricker} G.~R.,  et~al., 2015, \mn@doi [Journal of Astronomical Telescopes,
  Instruments, and Systems] {10.1117/1.JATIS.1.1.014003}, \href
  {https://ui.adsabs.harvard.edu/abs/2015JATIS...1a4003R} {1, 014003}

\bibitem[\protect\citeauthoryear{{Romanova}, {Ustyugova}, {Koldoba}  \&
  {Lovelace}}{{Romanova} et~al.}{2013}]{Romanova2013}
{Romanova} M.~M.,  {Ustyugova} G.~V.,  {Koldoba} A.~V.,   {Lovelace} R.~V.~E.,
  2013, \mn@doi [\mnras] {10.1093/mnras/sts670}, \href
  {https://ui.adsabs.harvard.edu/abs/2013MNRAS.430..699R} {430, 699}

\bibitem[\protect\citeauthoryear{{Storey} \& {Zeippen}}{{Storey} \&
  {Zeippen}}{2000}]{StoreyZeippen2000}
{Storey} P.~J.,  {Zeippen} C.~J.,  2000, \mn@doi [\mnras]
  {10.1046/j.1365-8711.2000.03184.x}, \href
  {https://ui.adsabs.harvard.edu/abs/2000MNRAS.312..813S} {312, 813}

\bibitem[\protect\citeauthoryear{{Tonry} et~al.,}{{Tonry}
  et~al.}{2012}]{Tonry2012}
{Tonry} J.~L.,  et~al., 2012, \mn@doi [\apj] {10.1088/0004-637X/750/2/99},
  \href {http://adsabs.harvard.edu/abs/2012ApJ...750...99T} {750, 99}

\bibitem[\protect\citeauthoryear{{Weise}, {Launhardt}, {Setiawan}  \&
  {Henning}}{{Weise} et~al.}{2010}]{Weise2010}
{Weise} P.,  {Launhardt} R.,  {Setiawan} J.,   {Henning} T.,  2010, \mn@doi
  [\aap] {10.1051/0004-6361/201014453}, \href
  {https://ui.adsabs.harvard.edu/abs/2010A&A...517A..88W} {517, A88}

\bibitem[\protect\citeauthoryear{{Wright} et~al.,}{{Wright}
  et~al.}{2010}]{wise_ref}
{Wright} E.~L.,  et~al., 2010, \mn@doi [\aj] {10.1088/0004-6256/140/6/1868},
  \href {http://adsabs.harvard.edu/abs/2010AJ....140.1868W} {140, 1868}

\bibitem[\protect\citeauthoryear{{Zacharias}, {Finch}, {Girard}, {Henden},
  {Bartlett}, {Monet}  \& {Zacharias}}{{Zacharias} et~al.}{2013}]{ucac4}
{Zacharias} N.,  {Finch} C.~T.,  {Girard} T.~M.,  {Henden} A.,  {Bartlett}
  J.~L.,  {Monet} D.~G.,   {Zacharias} M.~I.,  2013, \mn@doi [\aj]
  {10.1088/0004-6256/145/2/44}, \href
  {http://adsabs.harvard.edu/abs/2013AJ....145...44Z} {145, 44}

\makeatother
\end{thebibliography}



\appendix

\section{Photometry for V555 Ori}

\begin{table*}
\centering
\caption{Photometry data measured with the Schmidt telescope.}
\label{schmidt_magnitudes}
\begin{tabular}[b]{cccccccccc}
\hline
Date & MJD  & {\it B}& {\it V}&{\it R$\mathrm{\sb{c}}$}&{\it I$\mathrm{\sb{c}}$}\\

\hline
2017 Jan 28 & 57781.88 & 18.07 $\pm$ 0.14 & 16.32 $\pm$ 0.05 & 15.62 $\pm$ 0.04 & 14.62 $\pm$ 0.03\\
2017 Feb 05 & 57789.78 & 17.47 $\pm$ 0.15 & 16.38 $\pm$ 0.10 & 15.50 $\pm$ 0.06 & 14.70 $\pm$ 0.08\\
2017 Feb 10 & 57794.85 & 17.73 $\pm$ 0.18 & 16.55 $\pm$ 0.10 & 15.70 $\pm$ 0.08 & 14.86 $\pm$ 0.07\\
2017 Mar 02 & 57814.81 & 17.97 $\pm$ 0.17 & 16.59 $\pm$ 0.11 & 15.85 $\pm$ 0.09 & 14.99 $\pm$ 0.08\\
2017 Oct 12 & 58038.10 & 16.54 $\pm$ 0.24 & 15.08 $\pm$ 0.14 & 14.07 $\pm$ 0.10 & 13.12 $\pm$ 0.08\\ 
2017 Oct 14 & 58040.06 & 17.73 $\pm$ 0.25 & 16.29 $\pm$ 0.12 & 15.25 $\pm$ 0.09 & 14.22 $\pm$ 0.09\\ 
2017 Oct 15 & 58041.07 & 17.62 $\pm$ 0.26 & 16.16 $\pm$ 0.15 & 15.12 $\pm$ 0.10 & 14.10 $\pm$ 0.09\\ 
2017 Oct 19 & 58045.10 & 18.18 $\pm$ 0.23 & 16.62 $\pm$ 0.15 & 15.47 $\pm$ 0.11 & 14.40 $\pm$ 0.10\\ 
2017 Nov 04 & 58061.02 & 16.49 $\pm$ 0.08 & 15.10 $\pm$ 0.04 & 14.23 $\pm$ 0.05 & 13.32 $\pm$ 0.01\\
2017 Nov 06 & 58063.99 & 18.27 $\pm$ 0.08 & 16.61 $\pm$ 0.03 & 15.59 $\pm$ 0.06 & 14.69 $\pm$ 0.02\\
2017 Nov 07 & 58064.98 & 19.41 $\pm$ 0.14 & 17.30 $\pm$ 0.04 & 16.10 $\pm$ 0.04 & 15.00 $\pm$ 0.03\\
2017 Nov 25 & 58082.02 & 17.82 $\pm$ 0.23 & 16.27 $\pm$ 0.10 & 15.25 $\pm$ 0.10 & 14.28 $\pm$ 0.04\\
2017 Nov 27 & 58084.97 & 17.99 $\pm$ 0.06 & 16.63 $\pm$ 0.02 & 15.71 $\pm$ 0.03 & 14.82 $\pm$ 0.01\\
2017 Dec 05 & 58092.04 & 17.69 $\pm$ 0.06 & 15.80 $\pm$ 0.03 & 14.86 $\pm$ 0.03 & 13.82 $\pm$ 0.01\\
2017 Dec 07 & 58094.91 & 17.78 $\pm$ 0.13 & 16.12 $\pm$ 0.05 & 15.30 $\pm$ 0.04 & 14.27 $\pm$ 0.02\\
2017 Dec 16 & 58103.95 & 17.24 $\pm$ 0.05 & 15.82 $\pm$ 0.03 & 14.81 $\pm$ 0.02 & 13.87 $\pm$ 0.02\\
2018 Jan 07 & 58125.88 & 17.79 $\pm$ 0.06 & 16.25 $\pm$ 0.03 & 15.27 $\pm$ 0.05 & 14.23 $\pm$ 0.02\\
2018 Jan 14 & 58132.91 & 17.79 $\pm$ 0.09 & 16.06 $\pm$ 0.01 & 15.20 $\pm$ 0.01 & 14.12 $\pm$ 0.01\\
2018 Jan 25 & 58143.82 & 17.39 $\pm$ 0.11 & 16.00 $\pm$ 0.01 & 15.12 $\pm$ 0.01 & 14.02 $\pm$ 0.01\\
2018 Feb 18 & 58167.79 & 17.86 $\pm$ 0.09 & 16.39 $\pm$ 0.03 & 15.58 $\pm$ 0.05 & 14.63 $\pm$ 0.01\\
2018 Feb 25 & 58174.80 & 18.15 $\pm$ 0.05 & 16.16 $\pm$ 0.03 & 15.46 $\pm$ 0.04 & 14.61 $\pm$ 0.02\\
2018 Sep 25 & 58386.06 & 18.36 $\pm$ 0.15 & 16.82 $\pm$ 0.04 & 15.84 $\pm$ 0.01 & 14.92 $\pm$ 0.01\\
2018 Sep 26 & 58387.10 & 17.68 $\pm$ 0.07 & 16.18 $\pm$ 0.02 & 15.40 $\pm$ 0.03 & 14.36 $\pm$ 0.01\\
2018 Sep 28 & 58389.08 & 18.30 $\pm$ 0.05 & 16.73 $\pm$ 0.02 & 15.93 $\pm$ 0.02 & 14.91 $\pm$ 0.02\\
2018 Sep 30 & 58391.10 & 17.73 $\pm$ 0.10 & 16.12 $\pm$ 0.02 & 15.33 $\pm$ 0.03 & 14.34 $\pm$ 0.01\\ 
2018 Oct 04 & 58395.10 & 18.49 $\pm$ 0.04 & 17.26 $\pm$ 0.01 & 16.45 $\pm$ 0.02 & 15.63 $\pm$ 0.02\\
2018 Oct 05 & 58396.13 & 18.27 $\pm$ 0.06 & 17.06 $\pm$ 0.03 & 16.26 $\pm$ 0.06 & 15.49 $\pm$ 0.02\\
2018 Oct 06 & 58397.12 & 18.04 $\pm$ 0.09 & 16.51 $\pm$ 0.02 & 15.61 $\pm$ 0.02 & 14.51 $\pm$ 0.02\\
2018 Oct 09 & 58400.13 & 18.83 $\pm$ 0.09 & 17.44 $\pm$ 0.01 & 16.65 $\pm$ 0.01 & 15.81 $\pm$ 0.01\\
2018 Oct 10 & 58401.14 & 18.30 $\pm$ 0.06 & 16.93 $\pm$ 0.01 & 16.07 $\pm$ 0.02 & 15.10 $\pm$ 0.01\\
2018 Oct 25 & 58416.09 & 19.02 $\pm$ 0.14 & 17.04 $\pm$ 0.01 & 16.03 $\pm$ 0.01 & 14.98 $\pm$ 0.01\\
2019 Feb 22 & 58536.83& 19.36 $\pm$ 0.11& 17.45 $\pm$ 0.09& 16.50 $\pm$ 0.08 & 15.46 $\pm$ 0.05 \\
2019 Dec 26 & 58843.98& 17.22 $\pm$ 0.08& 15.60 $\pm$ 0.07& 14.73 $\pm$ 0.06 & 13.72 $\pm$ 0.04 \\
2019 Dec 29 & 58846.88& 18.22 $\pm$ 0.04& 16.55 $\pm$ 0.07& 15.69 $\pm$ 0.06 & 14.63 $\pm$ 0.04 \\
2020 Jan 05 & 58853.94& 17.38 $\pm$ 0.17& 16.08 $\pm$ 0.11& 15.15 $\pm$ 0.12 & 14.11 $\pm$ 0.06 \\
\hline
\end{tabular}
\end{table*}

\begin{table*}
\caption{Photometry data measured with the SMARTS telescope.}
\label{phot_tab}
\begin{tabular}{ccccccccr}
\hline
Date        & MJD      & {\it V}&{\it R$\mathrm{\sb{c}}$}&{\it I$\mathrm{\sb{c}}$}& {\it J}&{\it H}&{\it K$\mathrm{\sb{S}}$}\\
\hline
2017 Dec 07 & 58094.23 & 16.72 $\pm$ 0.01 & 15.74 $\pm$ 0.03 & 14.76 $\pm$ 0.01 & 12.365 $\pm$ 0.02 & 10.959 $\pm$ 0.02 & 9.905 $\pm$ 0.02 \\
2017 Dec 10 & 58097.22 & 15.86 $\pm$ 0.02 & 14.87 $\pm$ 0.04 & 13.84 $\pm$ 0.02 & 11.767 $\pm$ 0.02 & 10.568 $\pm$ 0.02 & 9.755 $\pm$ 0.03 \\
2017 Dec 12 & 58099.26 & 16.12 $\pm$ 0.05 & 15.19 $\pm$ 0.06 & 14.39 $\pm$ 0.09 & 12.095 $\pm$ 0.02 & 10.820 $\pm$ 0.02 & 9.884 $\pm$ 0.02 \\
2017 Dec 14 & 58101.22 & 15.80 $\pm$ 0.02 & 14.86 $\pm$ 0.03 & 13.88 $\pm$ 0.01 & 11.830 $\pm$ 0.02 & 10.642 $\pm$ 0.02 & 9.789 $\pm$ 0.02 \\
2017 Dec 16 & 58103.16 & 15.65 $\pm$ 0.02 & 14.74 $\pm$ 0.02 & 13.81 $\pm$ 0.01 & 11.817 $\pm$ 0.02 & 10.622 $\pm$ 0.02 & 9.763 $\pm$ 0.03 \\
2017 Dec 17 & 58104.23 & 15.80 $\pm$ 0.02 & 14.92 $\pm$ 0.03 & 14.03 $\pm$ 0.01 & 12.027 $\pm$ 0.02 & 10.785 $\pm$ 0.02 & 9.900 $\pm$ 0.02 \\
2017 Dec 19 & 58106.22 & 15.54 $\pm$ 0.03 & 14.65 $\pm$ 0.02 & 13.70 $\pm$ 0.02 & 11.650 $\pm$ 0.02 & 10.503 $\pm$ 0.02 & 9.702 $\pm$ 0.02 \\
2017 Dec 21 & 58108.17 & 16.45 $\pm$ 0.02 & 15.50 $\pm$ 0.04 & 14.57 $\pm$ 0.01 & 12.329 $\pm$ 0.02 & 10.969 $\pm$ 0.02 & 9.965 $\pm$ 0.02 \\
2017 Dec 23 & 58110.21 & 15.66 $\pm$ 0.02 & 14.72 $\pm$ 0.04 & 13.80 $\pm$ 0.01 & 11.794 $\pm$ 0.02 & 10.648 $\pm$ 0.02 & 9.825 $\pm$ 0.02 \\
2017 Dec 25 & 58112.26 & 16.44 $\pm$ 0.03 & 15.44 $\pm$ 0.03 & 14.50 $\pm$ 0.02 & 12.165 $\pm$ 0.02 & 10.842 $\pm$ 0.02 & 9.920 $\pm$ 0.02 \\
2017 Dec 27 & 58114.22 & 16.46 $\pm$ 0.03 & 15.60 $\pm$ 0.04 & 14.67 $\pm$ 0.01 & 12.354 $\pm$ 0.02 & 10.972 $\pm$ 0.02 & 9.995 $\pm$ 0.02 \\
2017 Dec 29 & 58116.18 & 15.82 $\pm$ 0.02 & 14.86 $\pm$ 0.02 & 13.90 $\pm$ 0.01 & 11.821 $\pm$ 0.02 & 10.623 $\pm$ 0.02 & 9.775 $\pm$ 0.02 \\
2017 Dec 31 & 58118.19 & 16.25 $\pm$ 0.03 & 15.42 $\pm$ 0.05 & 14.40 $\pm$ 0.03 & 12.303 $\pm$ 0.02 & 10.945 $\pm$ 0.03 & 9.978 $\pm$ 0.03 \\
2018 Jan 06 & 58124.21 & 15.75 $\pm$ 0.02 & 14.76 $\pm$ 0.02 & 13.78 $\pm$ 0.01 & 11.712 $\pm$ 0.02 & 10.556 $\pm$ 0.02 & 9.744 $\pm$ 0.02 \\
2018 Jan 08 & 58126.10 & 16.21 $\pm$ 0.03 & 15.28 $\pm$ 0.05 & 14.30 $\pm$ 0.02 & 12.040 $\pm$ 0.02 & 10.757 $\pm$ 0.02 & 9.821 $\pm$ 0.04 \\
2018 Feb 03 & 58152.12 & 15.86 $\pm$ 0.02 & 14.94 $\pm$ 0.06 & 13.95 $\pm$ 0.03 & 11.829 $\pm$ 0.02 & 10.659 $\pm$ 0.02 & 9.823 $\pm$ 0.03 \\
2018 Feb 04 & 58153.10 & 16.33 $\pm$ 0.01 & 15.40 $\pm$ 0.03 & 14.40 $\pm$ 0.01 & 12.180 $\pm$ 0.02 & 10.866 $\pm$ 0.02 & 9.963 $\pm$ 0.02 \\
2018 Feb 05 & 58154.08 & 16.35 $\pm$ 0.02 & 15.32 $\pm$ 0.03 & 14.30 $\pm$ 0.01 & 11.991 $\pm$ 0.02 & 10.746 $\pm$ 0.02 & 9.855 $\pm$ 0.02 \\
2018 Feb 06 & 58155.10 & 15.87 $\pm$ 0.02 & 14.92 $\pm$ 0.03 & 13.89 $\pm$ 0.01 & 11.821 $\pm$ 0.02 & 10.638 $\pm$ 0.02 & 9.823 $\pm$ 0.02 \\
2018 Mar 03 & 58180.11 & 16.52 $\pm$ 0.02 & 15.63 $\pm$ 0.04 & 14.73 $\pm$ 0.01 & 12.358 $\pm$ 0.02 & 10.953 $\pm$ 0.02 & 9.985 $\pm$ 0.02 \\
2018 Mar 04 & 58181.06 & 16.61 $\pm$ 0.01 & 15.82 $\pm$ 0.02 & 15.04 $\pm$ 0.01 & 12.716 $\pm$ 0.02 & 11.188 $\pm$ 0.02 & 10.036 $\pm$ 0.02 \\
2018 Mar 05 & 58182.05 & 16.64 $\pm$ 0.03 & 15.87 $\pm$ 0.03 & 15.14 $\pm$ 0.01 & 12.806 $\pm$ 0.02 & 11.278 $\pm$ 0.02 & 10.106 $\pm$ 0.02 \\
2018 Mar 06 & 58183.04 & 16.52 $\pm$ 0.03 & 15.65 $\pm$ 0.06 & 14.79 $\pm$ 0.02 & 12.447 $\pm$ 0.02 & 11.021 $\pm$ 0.02 & 9.965 $\pm$ 0.02 \\
2018 Apr 03 & 58211.02 & 16.32 $\pm$ 0.03 & 15.50 $\pm$ 0.05 & 14.65 $\pm$ 0.01 & 12.291 $\pm$ 0.02 & 10.878 $\pm$ 0.02 & 9.793 $\pm$ 0.02 \\
2018 Apr 04 & 58212.00 & 15.98 $\pm$ 0.02 & 15.08 $\pm$ 0.03 & 14.11 $\pm$ 0.01 & 11.940 $\pm$ 0.02 & 10.664 $\pm$ 0.02 & 9.744 $\pm$ 0.02 \\
2018 Apr 05 & 58213.00 & 15.45 $\pm$ 0.07 & 14.56 $\pm$ 0.11 & 13.67 $\pm$ 0.03 & 11.678 $\pm$ 0.02 & 10.504 $\pm$ 0.02 & 9.664 $\pm$ 0.02 \\
2018 Apr 06 & 58214.00 & 16.12 $\pm$ 0.03 & 15.22 $\pm$ 0.04 & 14.20 $\pm$ 0.01 & 11.989 $\pm$ 0.02 & 10.720 $\pm$ 0.02 & 9.788 $\pm$ 0.02 \\
2018 Nov 12 & 58434.26 & 17.16 $\pm$ 0.02 & 16.28 $\pm$ 0.03 & 15.41 $\pm$ 0.01 & 12.990 $\pm$ 0.02 & 11.386 $\pm$ 0.02 & 10.129 $\pm$ 0.02 \\
2018 Nov 13 & 58435.26 & 16.97 $\pm$ 0.01 & 16.10 $\pm$ 0.04 & 15.19 $\pm$ 0.01 & 12.838 $\pm$ 0.02 & 11.272 $\pm$ 0.02 & 10.049 $\pm$ 0.02 \\
2018 Nov 14 & 58436.27 & 17.03 $\pm$ 0.01 & 16.07 $\pm$ 0.02 & 15.14 $\pm$ 0.01 & 12.754 $\pm$ 0.02 & 11.226 $\pm$ 0.02 & 10.034 $\pm$ 0.02 \\
2018 Nov 15 & 58437.25 & 17.31 $\pm$ 0.01 & 16.45 $\pm$ 0.03 & 15.61 $\pm$ 0.01 & 13.169 $\pm$ 0.02 & 11.512 $\pm$ 0.02 & 10.167 $\pm$ 0.02 \\
2018 Nov 16 & 58438.23 & 17.47 $\pm$ 0.01 & 16.53 $\pm$ 0.03 & 15.60 $\pm$ 0.01 & 13.137 $\pm$ 0.02 & 11.493 $\pm$ 0.02 & 10.192 $\pm$ 0.02 \\
2018 Nov 17 & 58439.25 & 17.25 $\pm$ 0.01 & 16.32 $\pm$ 0.03 & 15.40 $\pm$ 0.01 & 13.013 $\pm$ 0.02 & 11.416 $\pm$ 0.02 & 10.151 $\pm$ 0.02 \\
2018 Nov 18 & 58440.23 & 17.00 $\pm$ 0.02 & 16.12 $\pm$ 0.02 & 15.26 $\pm$ 0.01 & 13.085 $\pm$ 0.02 & 11.483 $\pm$ 0.02 & 10.186 $\pm$ 0.02 \\
2018 Nov 19 & 58441.27 & 17.02 $\pm$ 0.01 & 16.13 $\pm$ 0.02 & 15.26 $\pm$ 0.01 & 13.020 $\pm$ 0.02 & 11.425 $\pm$ 0.02 & 10.139 $\pm$ 0.02 \\
2018 Nov 20 & 58442.25 & 17.44 $\pm$ 0.01 & 16.53 $\pm$ 0.01 & 15.64 $\pm$ 0.01 & 13.313 $\pm$ 0.02 & 11.643 $\pm$ 0.02 & 10.259 $\pm$ 0.02 \\
2018 Nov 22 & 58444.21 & 17.34 $\pm$ 0.01 & 16.37 $\pm$ 0.01 & 15.38 $\pm$ 0.01 & 12.836 $\pm$ 0.02 & 11.275 $\pm$ 0.02 & 10.085 $\pm$ 0.03 \\
2018 Nov 24 & 58446.24 & 17.13 $\pm$ 0.01 & 16.13 $\pm$ 0.01 & 15.11 $\pm$ 0.01 & 12.578 $\pm$ 0.02 & 11.084 $\pm$ 0.02 & 9.962 $\pm$ 0.02 \\
2018 Nov 25 & 58447.26 & 17.37 $\pm$ 0.01 & 16.39 $\pm$ 0.02 & 15.40 $\pm$ 0.01 & 12.807 $\pm$ 0.02 & 11.248 $\pm$ 0.02 & 10.066 $\pm$ 0.02 \\
2019 Mar 02 & 58544.04 & 17.19 $\pm$ 0.02 & 16.20 $\pm$ 0.03 & 15.21 $\pm$ 0.01 & 12.542 $\pm$ 0.02 & 10.998 $\pm$ 0.02 & 9.827 $\pm$ 0.02 \\
2019 Mar 03 & 58545.04 & 17.32 $\pm$ 0.02 & 16.34 $\pm$ 0.03 & 15.37 $\pm$ 0.01 & 12.594 $\pm$ 0.02 & 11.029 $\pm$ 0.02 & 9.798 $\pm$ 0.02 \\
2019 Mar 04 & 58546.08 & 16.78 $\pm$ 0.02 & 15.68 $\pm$ 0.03 & 14.90 $\pm$ 0.01 & 12.323 $\pm$ 0.02 & 10.811 $\pm$ 0.02 & 9.663 $\pm$ 0.02 \\
2019 Mar 05 & 58547.04 & 16.22 $\pm$ 0.01 & 15.29 $\pm$ 0.02 & 14.3  $\pm$ 0.01 & 11.927 $\pm$ 0.02 & 10.593 $\pm$ 0.02 & 9.574 $\pm$ 0.02 \\
2019 Mar 06 & 58548.06 & 17.09 $\pm$ 0.04 & 16.21 $\pm$ 0.06 & 15.29 $\pm$ 0.02 & 12.623 $\pm$ 0.02 & 11.048 $\pm$ 0.02 & 9.825 $\pm$ 0.02 \\
2019 Mar 07 & 58549.07 & 17.41 $\pm$ 0.05 & 16.54 $\pm$ 0.05 & 15.73 $\pm$ 0.02 & 12.955 $\pm$ 0.02 & 11.270 $\pm$ 0.02 & 9.963 $\pm$ 0.02 \\
2019 Mar 08 & 58550.05 & 17.07 $\pm$ 0.02 & 16.12 $\pm$ 0.02 & 15.2  $\pm$ 0.01 & 12.440 $\pm$ 0.02 & 10.882 $\pm$ 0.02 & 9.721 $\pm$ 0.02 \\
2019 Mar 09 & 58551.04 & 16.80 $\pm$ 0.03 & 15.97 $\pm$ 0.03 & 15.15 $\pm$ 0.01 & 12.469 $\pm$ 0.02 & 10.901 $\pm$ 0.02 & 9.697 $\pm$ 0.02 \\
2019 Mar 10 & 58552.04 & 16.67 $\pm$ 0.01 & 15.79 $\pm$ 0.02 & 14.86 $\pm$ 0.01 & 12.281 $\pm$ 0.02 & 10.837 $\pm$ 0.02 & 9.716 $\pm$ 0.02 \\
2019 Mar 11 & 58553.04 & 17.17 $\pm$ 0.02 & 16.39 $\pm$ 0.04 & 15.65 $\pm$ 0.01 & 12.957 $\pm$ 0.02 & 11.271 $\pm$ 0.02 & 9.978 $\pm$ 0.02 \\
2019 Mar 12 & 58554.02 & 17.46 $\pm$ 0.01 & 16.56 $\pm$ 0.03 & 15.93 $\pm$ 0.01 & 13.234 $\pm$ 0.02 & 11.470 $\pm$ 0.02 & 10.079 $\pm$ 0.02 \\
2019 Mar 13 & 58555.05 & 17.17 $\pm$ 0.02 & 16.32 $\pm$ 0.03 & 15.55 $\pm$ 0.01 & 12.830 $\pm$ 0.02 & 11.200 $\pm$ 0.02 & 9.946 $\pm$ 0.02 \\
\hline
\end{tabular}
\end{table*}



\bsp	
\label{lastpage}
\end{document}